\documentclass[useAMS,usenatbib]{mn2e}
\usepackage{graphicx}
\usepackage{deluxetable}
\usepackage{cleveref}
\usepackage{amsmath}
\usepackage{amssymb}
\usepackage{float}
\usepackage{rotating}
\usepackage{empheq}
\usepackage{mathtools}
\usepackage{comment}
\usepackage{fixltx2e}
\usepackage{color}
\usepackage{soul}

\title{Investigation of the environment around close-in transiting exoplanets using \texttt{CLOUDY}}
\author[Turner et al.]
  {Jake D. Turner$^{1}$, Duncan Christie$^{1}$, Phil Arras$^{1}$, Robert E. Johnson$^{2}$, Carl Schmidt$^{1,2}$
 \\
  $^1$Department of Astronomy, University of Virginia, Charlottesville, VA 22904, USA\\
 $^2$Department of Materials Science $\&$ Engineering, University of Virginia, Charlottesville, VA 22904, USA
  }
\pagerange{\pageref{firstpage}--\pageref{lastpage}} \pubyear{2015}

\begin{document}
\label{firstpage}
\maketitle
\begin{abstract}

It has been suggested that hot stellar wind gas in a bow shock around an exoplanet is sufficiently opaque to absorb stellar photons and give rise to an observable transit depth at optical and UV wavelengths. In the first part of this paper, we use the \texttt{CLOUDY} plasma simulation code to model the absorption from X-ray to radio wavelengths by 1-D slabs of gas in coronal equilibrium with varying densities ($10^{4}-10^{8} \, {\rm cm^{-3}}$) and temperatures ($2000-10^{6} \ {\rm K}$) illuminated by a solar spectrum. For slabs at coronal temperatures ($10^{6} \ {\rm K}$) and densities even orders of magnitude larger than expected for the compressed stellar wind ($10^{4}-10^{5} \, {\rm cm^{-3}}$), we find optical depths orders of magnitude too small ($> 3\times10^{-7}$) to explain the $\sim3\%$ UV transit depths seen with Hubble. Using this result and our modeling of slabs with lower temperatures ($2000-10^4 {\rm K}$), the conclusion is that the UV transits of WASP-12b and HD 189733b are likely due to atoms originating in the planet, as the stellar wind is too highly ionized. A corollary of this result is that transport of neutral atoms from the denser planetary atmosphere outward must be a primary consideration when constructing physical models. In the second part of this paper, additional calculations using \texttt{CLOUDY} are carried out to model a slab of planetary gas in radiative and thermal equilibrium with the stellar radiation field. Promising sources of opacity from the X-ray to radio wavelengths are discussed, some of which are not yet observed. 

\end{abstract}

\section{Introduction} \label{intro}

Exoplanet transits observed at broadband optical wavelengths probe relatively dense gas in the planet's atmosphere (e.g. \citealt{Charbonneau2002}; \citealt{Charbonneau2007}). By contrast, radiative transitions for abundant species and with large cross sections may probe more rarefied regions at much larger radii (e.g. \citealt{Madjar2003}). Absorption well before or after the broadband optical transit implies optically thick absorbing gas at large distances from the planet (e.g. \citealt{Llama2011}; \citealt{Lai2010}). Recent observations of asymmetry in transit lightcurves (e.g. \citealt{Fossati2010b}; \citealt{Haswell2012}; \citealt{Jaffel2013}), specifically an early ingress, point toward an enhanced density of absorbers on the leading side of the planet.

There are several space-based observations of transit light curves exhibiting an early UV ingress. An early UV ingress on WASP-12b was observed based on 5 data points from the Cosmic Origins Spectrograph (\textit{COS}) onboard the \textit{Hubble Space Telescope} (HST) in the NUVA (253.9--258.0 nm) near-UV wavelength region (\citealt{Fossati2010b}). Additionally, \citet{Haswell2012} observed an early ingress on WASP-12b in the NUVC (278.9 -- 282.9 ${\rm nm}$) near-UV wavelength region caused by Mg II and the NUVA region using an additional 5 data points from \textit{COS} on \textit{HST}. Sequential \textit{HST COS} observations of WASP-12b by \citet{Nichols2015} combined the data from the NUVA, NUVB (265.5 -- 271.1 ${\rm nm}$), and NUVC wavelength regions and also found an early ingress. The data from \citet{Haswell2012} and \citet{Nichols2015}, however, are noisy. They observed absorption before the start of the optical transit but never observed a pre-transit baseline. Interpretations of the observations become more difficult if you include all the data points and do not exclude certain data points (see fig 5 in \citealt{Haswell2012}). Follow-up observations of WASP-12b using the Space Telescope Imaging Spectrograph instrument (wavelength range from 290.0--570.0 ${\rm nm}$) on \textit{HST} \citep{Sing2013} resulted in a non-detection of a timing difference. There was also a reported early ingress on HD 189733b in the C II 133.5 ${\rm nm}$ line using \textit{COS} on \textit{HST} \citep{Jaffel2013}. However, these authors call for additional observations and caution that the detection could have been caused by unknown systematics.\\
\indent There are two main models for explaining the origin of an early ingress -- magnetic (\citealt{Vidotto2010,Vidotto2011a,Vidotto2011b}; \citealt{Vidotto2011c}; \citealt{Llama2011,llama2013}) and non-magnetic (\citealt{Lai2010}; \citealt{Bisikalo2013}; \citealt{Bisikalo2013A}) -- which we will now describe. A planet traveling supersonically through the ambient coronal medium will form a bow shock. An early ingress can result if the bow shock is leading the planet and the compressed post-shock material in the magnetosheath becomes
sufficiently opaque, absorbing the background starlight. The transit depth will then become a function of the geometry of the post-shock layer and the chemistry therein (\citealt{Llama2011,llama2013}). Ignoring thermal and ram pressure, the approximate location of the magnetopause\footnote{The approximate location of the magnetopause ($r_{m}$) can be found using the following equation $r_{m} = R_{p} \left( \frac{B_{p}}{B_{*}} \right)^{1/3} \frac{a}{R_{*}}$, where $R_{p}$ is the radius of the planet, $B_{p}$ is the planetary magnetic field strength at the surface of the planet, $B_{*}$ is the host star's magnetic field strength at the surface of the star, $a$ is the semi-major axis of the planet, and $R_{*}$ is the radius of the star (\citealt{Vidotto2011a}). Both the star and the planet magnetic field geometry are approximated as dipolar.} is found assuming pressure balance between the planetary magnetic field and the stellar wind magnetic field (\citealt{Vidotto2011a}). This type of bow shock will be referred as a magnetic bow shock in the rest of the paper. Thus, assuming that the stellar magnetic field is known, that the thermal and ram pressures are negligible, and that the stellar wind gas is opaque enough to cause detectable absorption, constraints on the planet's magnetic field can in principle be made by observing differences between ingress times in different wavelengths. The reason this effect is thought to occur in the UV is not entirely understood and is one area of investigation for this paper. \citet[hereafter VJH11a]{Vidotto2011a} presented a magnetic bow shock model that they applied to all transiting exoplanets and predicted that UV ingress asymmetries should be common in transiting exoplanets and tabulated a list of the 92 targets that should exhibit this effect.

A model ignoring magnetic fields has recently been proposed to explain the WASP-12b \textit{HST} observations (\citealt{Bisikalo2013}; \citealt{Bisikalo2013A}; see also \citealt{Lai2010}). These authors perform hydrodynamic simulations in 3-D and looked at the interaction between the stellar wind and the escaping planetary atmosphere. In order to find the stand off distance from the planet, the stellar wind ram pressure was balanced by pressure from the planetary gas assuming negligible contribution from the planetary and the stellar wind magnetic field (\citealt{Bisikalo2013}). This type of bow shock will be referred to as a non-magnetic shock in the paper. Gas leaving the L1 point is bent forward in the orbit by the Coriolis force, leading to a distribution of gas leading the planet. As with the VJH11a model, the exact absorbing species in the non-magnetic bow shock that can cause an early ingress is unknown and is also investigated in this paper.

While both the magnetic bow shock theory and initial observations focused on the UV, a recent study by \citealt{Cauley2015} opens up the possibility of applying the VJH11a magnetic bow shock model to other wavelengths. \citealt{Cauley2015} observe an asymmetric transit in HD 189733b using \textit{HiRES} on \textit{Keck I} in the hydrogen Balmer lines. These authors apply the magnetic bow shock model to their observations by balancing the planetary magnetic field pressure with the stellar wind pressure assuming negligible contribution from the stellar magnetic field.

Furthermore, there are extensive near-UV observations from the ground looking for asymmetries due to bow shocks. Absorption is thought to be observable in ground-based observations because the spectral region covered by the ground-based near-UV observations and at least one of the \textit{HST} NUVA, NUVB, or NUVC filters include strong resonance lines (e.g. Ne I, Na I, Mg I, Al I, Mn I, Fe I, Co I, Ni I, Cu I; \citealt{Morton1991,Morton2000}; \citealt{Sansonetti2005}) and lots of lines from ionised abundant elements (e.g. Ne III, Na II, Cl II, Ca II, Ca III, Sc II, V II, Cr II, Mn II, Fe II, Co II, Ni I, Cu II, Cu III; \citealt{Morton1991,Morton2000}; \citealt{Sansonetti2005}). However, near-UV ground-based broad-band (303--415 ${\rm nm}$) photometry observations of 19 exoplanets (including WASP-12b) have all resulted in non-detections (\citealt{Southworth2012b}; \citealt{Pearson2014}; \citealt{Turner2013a}; \citealt{Bento2014}; \citealt{Copperwheat2013}; \citealt{Zellem2015}; Turner et al. in press) despite the prediction by \citet{Copperwheat2013} that an early ingress could be seen in these filters. The ground-based broad-band photometry non-detections might imply that the coronal material does not absorb sufficiently strongly at the observed wavelengths, the absorption of specific spectral lines being diluted by using broad-band filters, or one of the requirements presented by VJH11a for a stable bow shock detection not being satisfied (\citealt{Turner2013a}; \citealt{Pearson2014}; \citealt{Vidotto2015}; Turner et al. in press). Our study will shed light onto the correct interpretation of the ground-based observations. 

In this study we investigate all known UV opacity sources by performing a detailed radiative transfer, chemical, and ionization analysis of the coronal plasma environment in the bow shocks around close-in exoplanets using the plasma simulation code \texttt{CLOUDY} \citep{Ferland2013}. This modeling will test the basic assumption of VJH11a that the stellar wind is sufficiently opaque to cause the early transit ingress. We do not, however, consider the opacity from a population of energetic atoms which arise due to charge exchange reactions between ``cold" atoms from the planet and ``hot" stellar wind protons \citep{2008Natur.451..970H, Tremblin2013}. Additionally, we model with \texttt{CLOUDY} the planetary gas in thermal and ionization equilibrium with the stellar radiation field to determine what species may be observable. The overview of the \texttt{CLOUDY} modeling can be found in \S~\ref{sec:modeling} and the results and discussion for the coronal gas and planetary gas can be found in \S~\ref{sec:coronal} and \S~\ref{sec:planetary}, respectively.


 
\section{Overview of \texttt{CLOUDY}}  \label{sec:modeling}

\texttt{CLOUDY} is a widely-used plasma simulation code designed to simulate the ionization, chemical, and thermal state of an astronomical plasma exposed to an external radiation field and to predict its emission and absorption spectra (\citealt{Ferland2013}; \citealt{Ferland1998}). It accounts for the entire electromagnetic spectrum from hard X-rays to the radio and includes the 30 lightest elements in its calculations. By default, the abundances of these elements are assumed to be solar but this can be changed. \texttt{CLOUDY} self-consistently balances all ionization, excitation, and microphysical (e.g. inner shell ionization and charge exchange) rates for all constituents (\citealt{Badnell2003}; \citealt{Bryans2006}). For the ionic and molecular emission data, \texttt{CLOUDY} uses the \texttt{CHIANTI} database (version 7: \citealt{Dere1997}; \citealt{Landi2012}), the \texttt{LAMDA} database (\citealt{Sch2005}), and its own Atomic and Molecular Database (\texttt{STOUT}; \citealt{Lykins2015}). All known spectral lines are taken into account from these three databases and level energies are from NIST when available (\citealt{NIST_ASD}). A thorough description of \texttt{STOUT} can be found in \citet{Lykins2015}. The program has been thoroughly tested and can be used for number densities up to $10^{15} \ {\rm cm^{-3}}$ and temperatures from 3 K to $10^{10}\, {\rm K}$ (\citealt{Ferland2013}). Additionally, all \texttt{CLOUDY} modeling performed in this study is done in 1-D.


The two main choices for the equilibrium condition enforced by \texttt{CLOUDY} are thermal and coronal equilibrium. In coronal equilibrium, the temperature is fixed and the collisional ionization rate is set by the gas temperature and ionization potentials of all the constituents. Photoionization also occurs in coronal equilibrium if an external radiation field is specified. Heating and cooling are not necessarily in equilibrium for the coronal case. The level populations are then determined by balancing all the relevant processes. In thermal equilibrium, the temperature of the slab of gas (cloud) is determined by the balance of cooling and heating of the gas.

\section{Modeling of the coronal gas with \texttt{CLOUDY}}\label{sec:coronal}

For the \texttt{CLOUDY} modeling, we simulate the output spectrum and transit depth in the conditions of the shocked stellar coronal gas in the magnetosheath. The spectrum used is the net transmitted spectrum by summing the attenuated incident and diffuse continua and lines\footnote{This spectrum is from column 5 in the continuum command in \texttt{CLOUDY}}. 

\begin{figure}
\center
\includegraphics[width=1\linewidth]{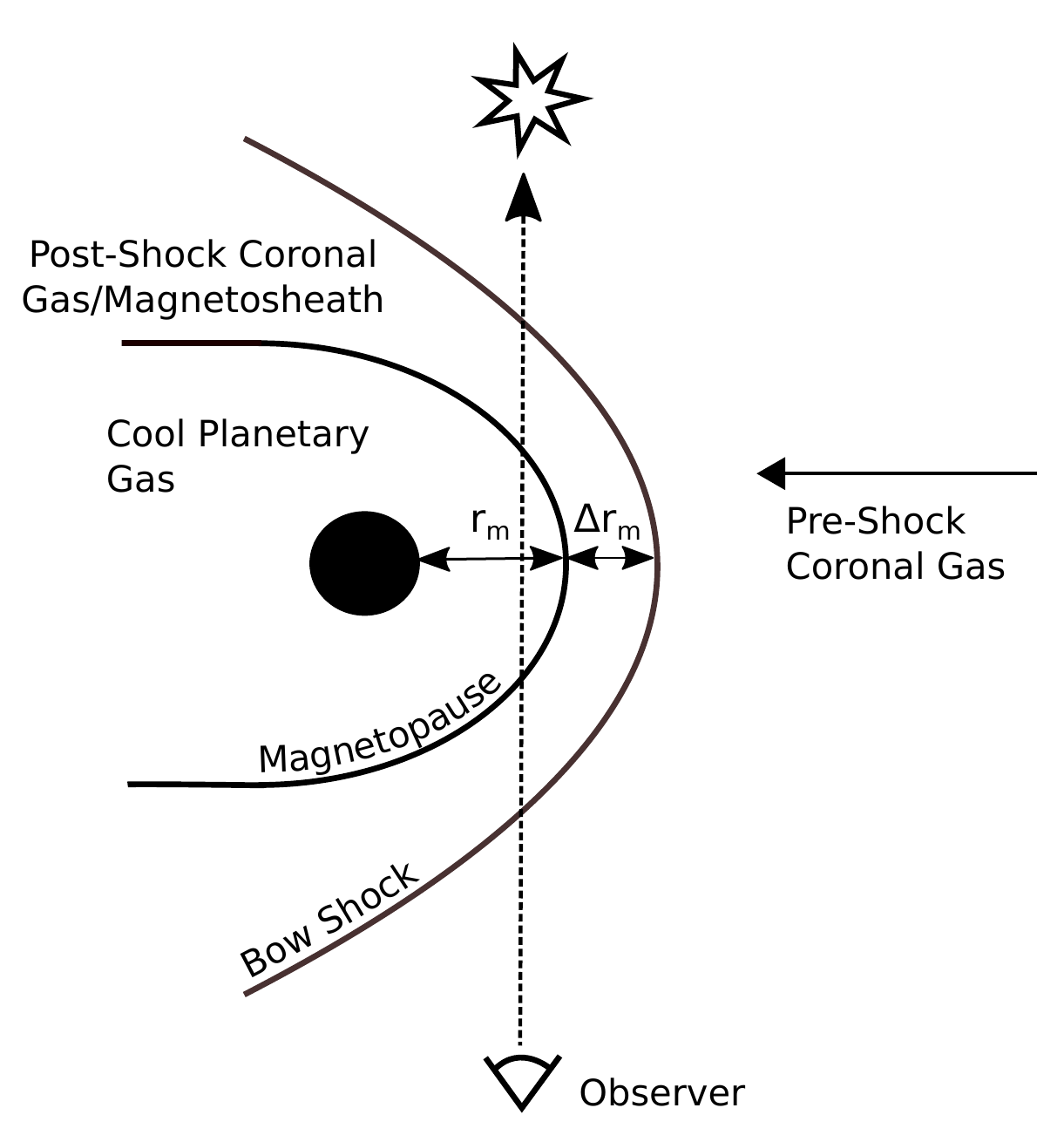}
\caption{Schematic of the magnetic bow shock geometry (not to scale). The stand-off distance, $r_{M}$, is determined by balancing the planetary magnetic field pressure with the ram pressure from the stellar wind. $\Delta r_{M}$ is the region of compressed stellar coronal material. This geometry also applies to non-magnetic bow shocks. }
\label{fig:bow_geo}
\end{figure}

\begin{figure}
\center
\includegraphics[width=1\linewidth]{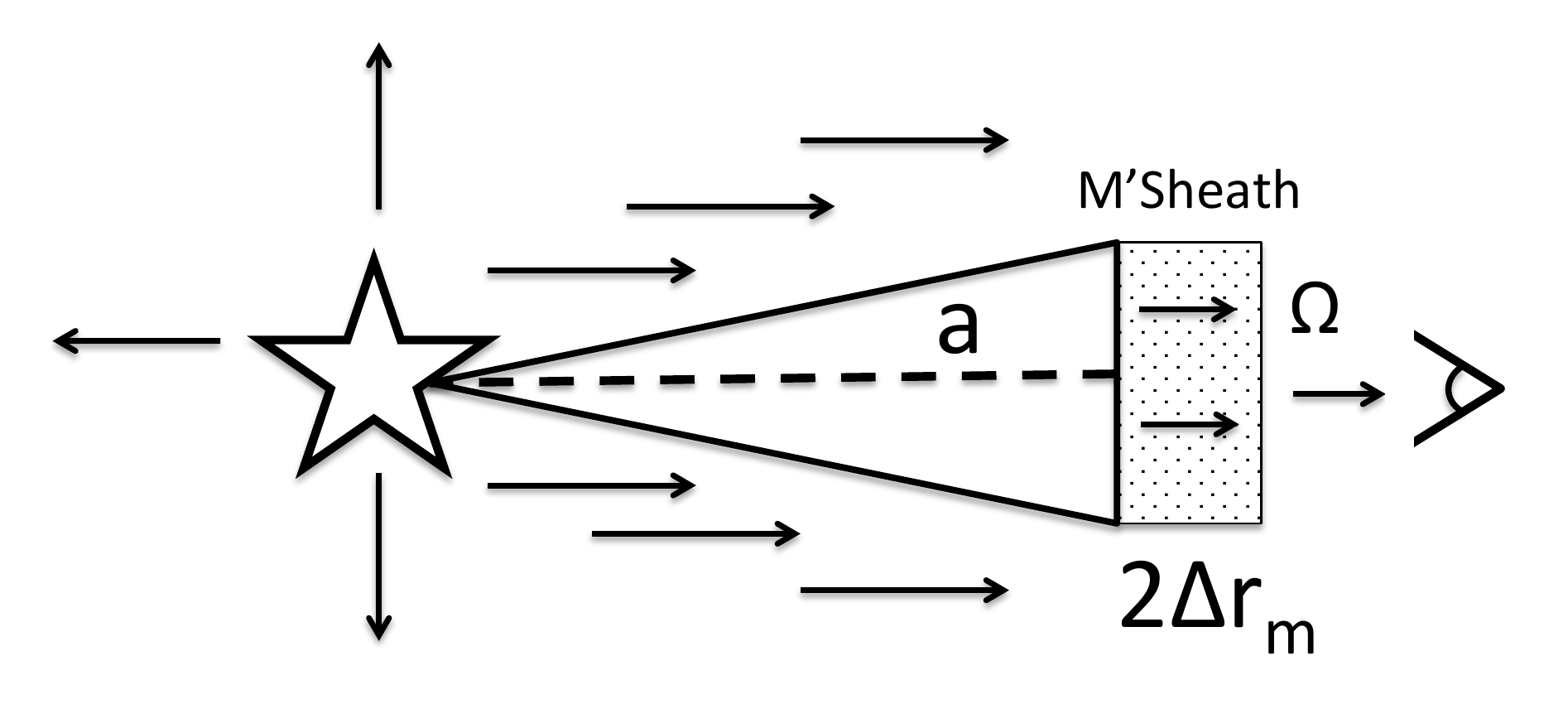}  \\
\caption{Schematic of the \texttt{CLOUDY} input model (not drawn to scale). The magnetosheath is modeled as a cloud with a width of 2$\Delta r_{m}$ and a covering factor equal to $\Omega/4\pi$. See $\S$ \ref{sec:coronal} for more details. The incident radiation strikes the cloud at a normal illumination.
}
\label{fig:cloudy_geo}
\end{figure}

A brief description of the geometry of the bow shock (Figure \ref{fig:bow_geo}) and the simplified geometry used for the \texttt{CLOUDY} modeling (Figure \ref{fig:cloudy_geo}) can be found below. A bow shock will form if the relative velocity between the planetary and the stellar coronal material is supersonic and will lead the exoplanet if the star's rotational period is greater than planet's orbital period (as is the case for most close-in exoplanets; VJ11a). The upstream region is composed of three regions of interest: the bow shock, the magnetosheath, and the magnetopause. In this paper we adopt magnetospheric terminology for the shock geometry; however, this geometry applies generally to any planet with a bow shock not just magnetic ones. The magnetosheath is the area of compressed coronal material and has a width of $\Delta r_{m}$. The magnetopause is the boundary between the magnetosphere and surrounding plasma from the stellar wind. Inside the magnetopause, the plasma is assumed to be composed predominantly of gas from the planet (See \S \ref{sec:planetary_stellar}). The stand-off distance, $r_{M}$, is determined by balancing the planetary magnetic field pressure with the ram pressure of the stellar wind. For the \texttt{CLOUDY} modeling, we model the magnetosheath as a cloud with a width of 2$\Delta r_{m}$ and a covering factor equal to $\Omega/4\pi$ (Figure \ref{fig:cloudy_geo}). This approximation to the cloud width is chosen as a simplification of the 3-D structure of the bow shock since photons along the line of sight go through the magnetosheath twice. The covering factor is the fraction of $4\pi$ sr covered by the cloud, as viewed from the star (the central source of radiation) and represents the fraction of the radiation field emitted by the star that actually strikes nebular gas. The line luminosities and intensities are the main characteristics that depend on the covering factor. The solid angle of the cloud as viewed from the central star is
\begin{align}
\Omega = \pi\frac{(r_{m} + \Delta r_{m})^2 }{2a^2}, \label{eq:omega}
\end{align}
where $r_{m}$ is the magnetospheric radius. Using Equation \ref{eq:omega}, we find a covering factor of 0.0011 for $r_{m} = 4.4 \ {\rm R_{p}}$, assuming a planetary magnetic field ($B_{p}$) of 4 G, and 0.0034 for $r_{m} = 8.0 \ {\rm R_{p}}$, assuming a $B_{p}$ of 30 G \citep{Lai2010}. Since the strength of the planetary magnetic fields are not known, we use a covering factor of 0.0034 in order to be conservative in testing weather the total absorption is substantial enough to explain the observations.


\begin{figure}
\center
\includegraphics[width=1\linewidth]{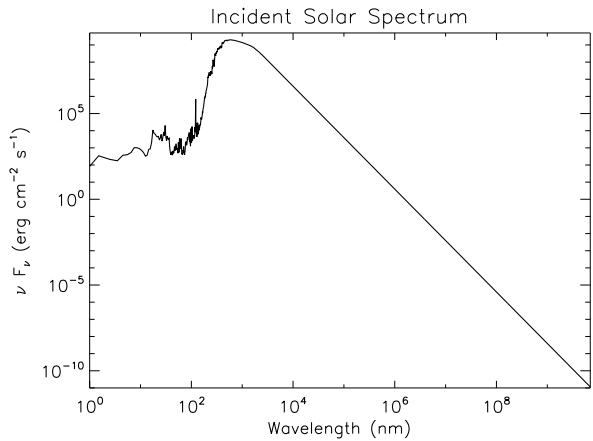} 
\caption{Solar spectrum input into \texttt{CLOUDY} as the external radiation field incident on the cloud. The spectrum was taken from stitching together data from the TIMED-SEE \citep{Woods2000} and SORCE \citep{Anderson2005} satellites.}
\label{fig:solar}
\end{figure}

We ran \texttt{CLOUDY}\footnote{Calculations were performed with version 13.00 of \texttt{CLOUDY} \citep{Ferland2013}} in an open geometry and in coronal equilibrium assuming a constant temperature of $1\times10^{6}\, {\rm K}$ (see description below about why this is a good assumption), a nominal magnetosheath width ($\Delta r_{m}$) of $0.24\, {\rm R_{Jup}}$ \citep{Llama2011}, an orbital distance ($a$) of $0.023\, {\rm au}$ (the orbital distance of WASP-12b; \citealt{Hebb2009}), and a solar metallicity ($[Fe/H]$) for the cloud (Table \ref{tb:CLOUDY_modelsetup}). The external radiation field was a solar spectrum taken from stitching together data from the TIMED-SEE \citep{Woods2000} and SORCE \citep{Anderson2005} satellites (Figure \ref{fig:solar}). We then varied the hydrogen density ($n_{H}$) in the cloud between $10^{4}-10^{8} \, {\rm cm^{-3}}$ and calculated the output spectrum. In order to correctly interpret the \texttt{CLOUDY} results it is important to note that $n_H$ is the density of hydrogen nuclei, irrespective of their form (protons, bound atoms, molecules). All the elements are scaled relative to $n_{H}$ based on their solar abundances.  

We explain below our choice of the parameter range explored for the hydrogen densities. The expected stellar wind density at the planet may be estimated as $n_{\rm H} \simeq 10\ {\rm cm^{-3}}(1\ {\rm AU}/a)^2 = 10^4\ {\rm cm^{-3}}\ (0.03\ {\rm AU}/a)^2$ by scaling the density of the solar wind as measured near Earth as $1/a^2$ with separation $a$ from the star (this profile is an approximation to the actual stellar wind profile measured for the sun; \citealt{McKenzie1997}). Hence the lower limit of the density range explored is thought to be the actual value around a solar-type star. The upper limit $n_{\rm H} \simeq 10^{8} \ {\rm cm^{-3}}$ is an estimate of the density at the base of the solar corona \citep{Withbroe1988}, the highest density expected for coronal gas. Since the base of the corona is at $a \sim 1.001 R_\star$, and the planets considered here orbit at $a \sim 3-9\ R_\star$, the planets are expected to be orbiting in an environment with density much smaller than the coronal base density. These densities are consistent with those found by \citet{Vidotto2010} and VJH11a for an hydrostatic isothermal corona (see table 1 in VJH11a). The actual stellar wind density at WASP-12b and HD189733b are not known but should be within the limits explored in this study. Assuming the Rankine-Hugoniot conditions for a $\gamma=5/3$ gas, we only expect a maximum increase of 4 in the plasma density across the bow shock between the stellar wind and magnetosheath. 

\begin{table}
\caption{\texttt{CLOUDY} model parameters }
\begin{tabular}{cc}
\hline
\multicolumn{1}{c}{Parameter} &
\multicolumn{1}{c}{Magnetosheath}\\
\hline
Hydrogen Density ($\rm cm^{-3}$) & $10^{4}-10^{8}$\\
Stellar Luminosity  ($L_{\sun}$)     &1 \\
Orbital Distance (au)   &$0.023$ \\
Magnetosheath Width (cm)   &$0.24\ R_{Jup}$\\
Shock Temperature (K)      &$1\times 10^{6}$\\
Metallicity          &Solar\\
Covering Factor ($\Omega/4\pi$)     &$0.0034$ \\
\hline
\end{tabular}
\label{tb:CLOUDY_modelsetup}
\end{table}

The temperature of the shock is one of the most important parameters in our modeling, therefore, we discuss our choice of shock temperature and the assumptions we use in greater detail below. A shock temperature is adopted that is comparable to that of the solar coronal gas ($T = 10^{6} {\rm K}$; \citealt{Aschwanden2005}) and subsequently we vary the specific choice of temperature in a parameter study (Section \ref{sec:results}). Previous works have posited that the post-shock gas, while initially hotter due to compression (increased by the square of the Mach number), undergoes radiative cooling to reach temperatures in the range of $10^4$ to $10^5\,{\rm K}$ \citep{Lai2010,Vidotto2010}. No attempts are made, however, to estimate whether this degree of cooling is realistic. To determine whether our adopted shock temperature is a reasonable assumption we compare the cooling rate of the gas to the dynamical time. \citet{Sutherland1993} find that for gas with temperatures on the order of $10^6- 10^{7}\,{\rm K}$, the cooling rate is approximately $10^{-22}n_{e}n \,{\rm erg \, cm^{3}\, s^{-1}}$ where $n$ is the total density of nuclei and $n_{e}$ is the number density of electrons.  The characteristic cooling time $t_{\rm cool}$ for gas with energy density $E$ and a cooling rate $\dot{E}$ is then
\begin{eqnarray}
t_{\rm cool} &\sim &\frac{E}{\dot{E}} \sim \frac{nkT}{\left(10^{-22}\,{\rm erg\, cm^{3}\,s^{-1}}\right)n_{\rm e}n} \\
 &\sim &10^8\left(\frac{T}{10^6\,{\rm K}}\right)\left(\frac{10^4\,{\rm cm^{-3}}}{n_{\rm e}}\right)\,\, {\rm s}\,\, . 
\end{eqnarray}
Adopting a thickness of the post-shock gas of $10^{10}\,{\rm cm}$, comparable to $R_{\rm p}$, and a characteristic velocity of $100\,{\rm km\, s^{-1}}$ we find a dynamical time $t_{\rm dyn}$ of
\begin{equation}
t_{\rm dyn} \sim \frac{10^{10}\, {\rm cm}}{100\,{\rm km\, s^{-1}}} = 10^3\, {\rm s}\,\, .
\end{equation}
The cooling time is thus much longer than the dynamical time, and the post-shock gas leading the planet should not be expected to cool significantly. Therefore, post-shock temperatures comparable to the coronal temperatures are reasonable within the context of the model being examined.

\begin{figure}
\center
\vspace{0.25cm}
\includegraphics[width=1\linewidth]{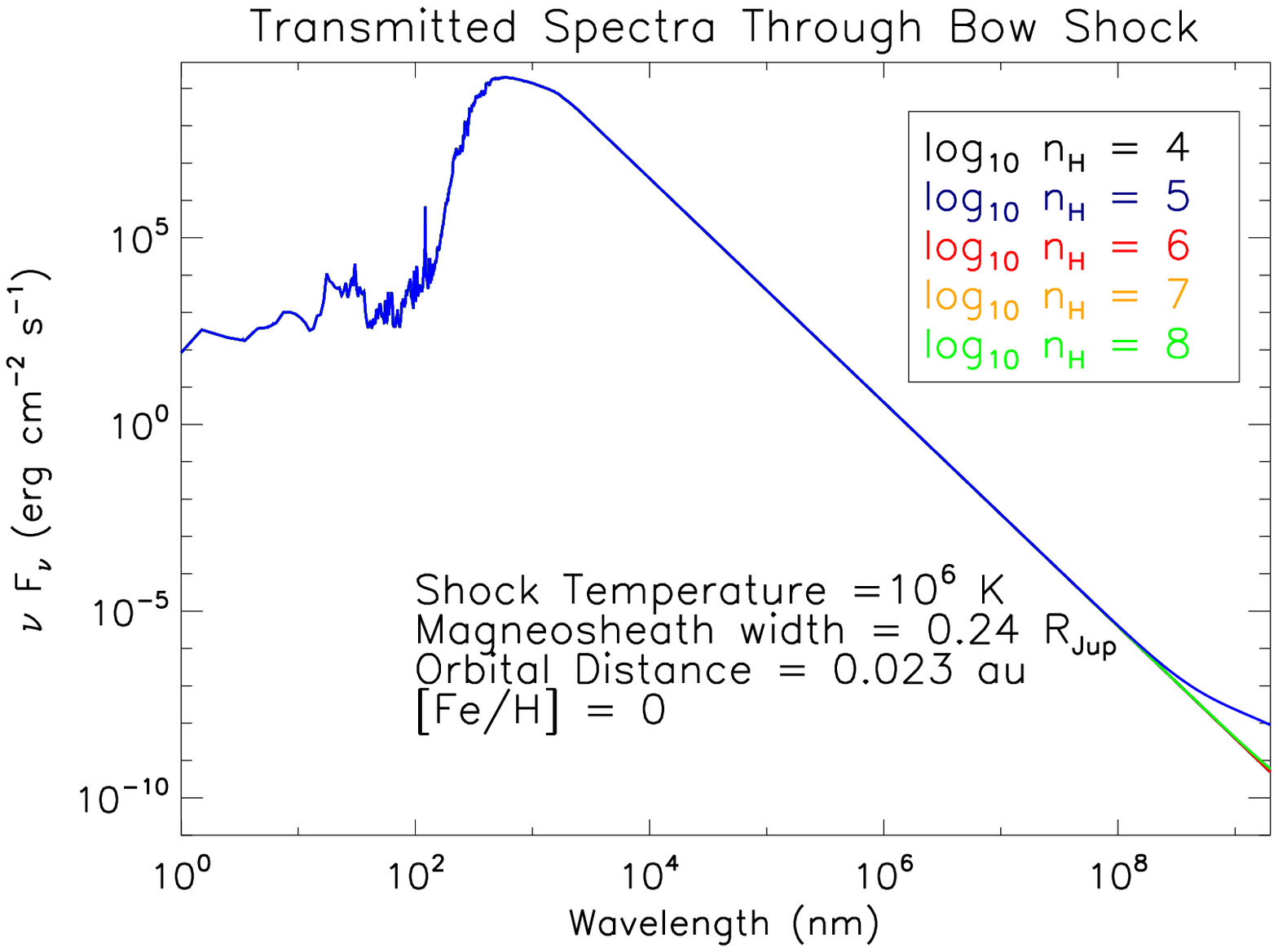}  \\
\includegraphics[width=1\linewidth]{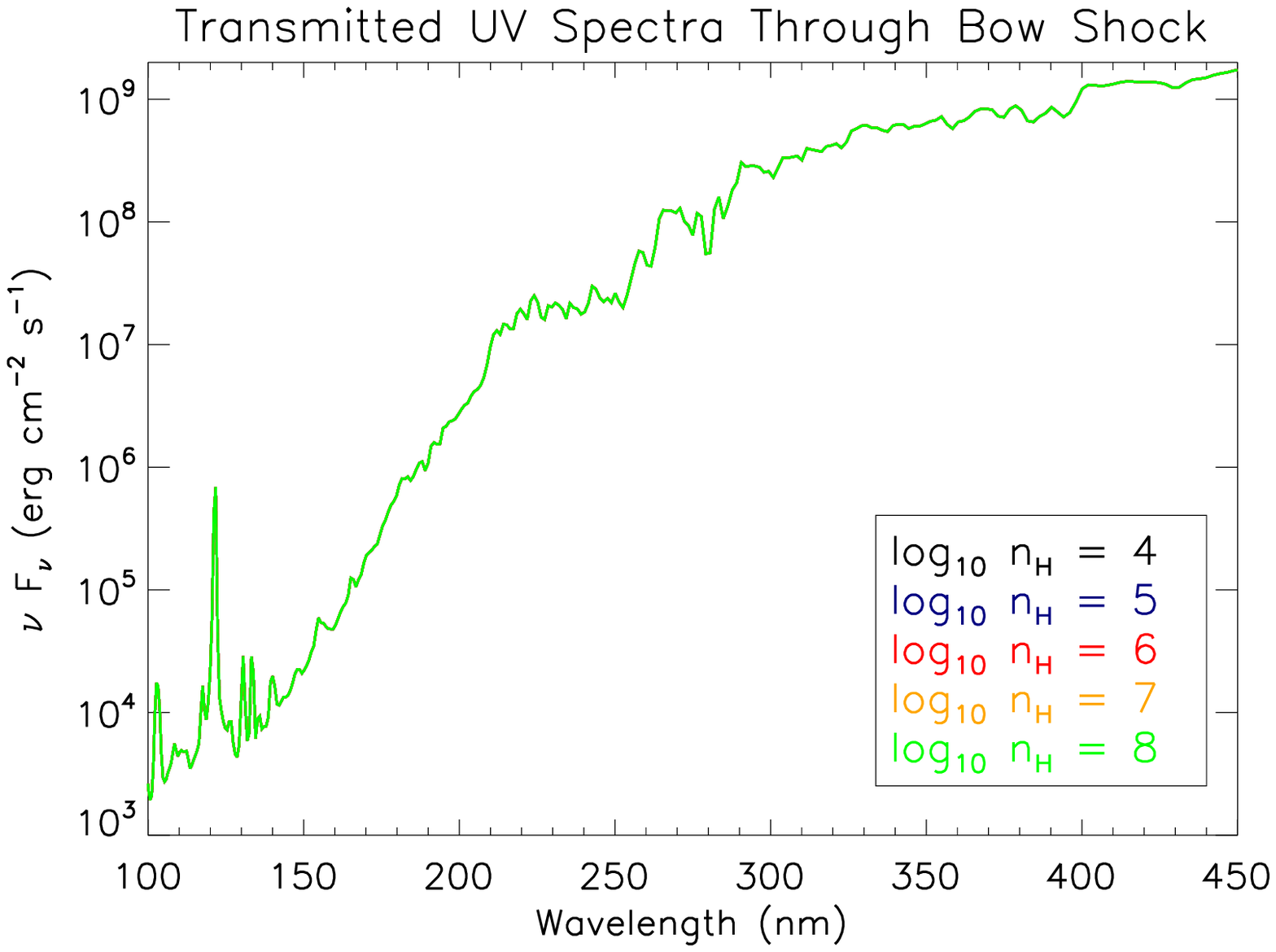}  \\
\caption{Output \texttt{CLOUDY} spectra of the magnetosheath with varying hydrogen densities ($n_{H}$) in the stellar corona for all (\textbf{Top}) and UV (\textbf{Bottom}) wavelengths. The output spectra are all identical to each other for the entire UV. 
}
\label{fig:spec}
\end{figure}

\subsection{Results} \label{sec:results}
The net transmitted spectrum of the shocked material is in Figure \ref{fig:spec} for all wavelengths and the UV wavelengths probed by \textit{HST} and the ground-based observations. Next, we calculate the transit depth, $\delta_{F}$, due to the magnetosheath being in front of the star during the transit assuming the magnetosheath is half of a spherical shell (see Figure \ref{fig:bow_geo}),  
\begin{align}
\delta_{F} = \frac{F_{bow}(\lambda) - F_{star}(\lambda)}{F_{Star}(\lambda)}\min\left( \frac{1}{2}\frac{ (r_{m} + \Delta r_{m})^2 }{R_{\sun}^2},1\right) \label{eq:deltaf}
\end{align}
where $F_{bow}$ is the flux from the bow shock calculated from the output of \texttt{CLOUDY}, $F_{star}$ is the flux from the star (the inputted solar spectrum), $r_{m}$ is the magnetospheric radius (set as $8 R_{Jup}$ to be conservative to produce the largest $\delta_{F}$; \citealt{Lai2010}), and $\Delta r_{m}$ is the width of the magnetosheath. The left term in Equation \ref{eq:deltaf} is essentially the opacity of the cloud. The geometric factor (equal to $\frac{1}{2}\frac{ (r_{m} + \Delta r_{m})^2 }{R_{\sun}^2}=0.336 = 33.6$\%$ $ for the nominal model) is applied since the output spectrum from \texttt{CLOUDY} does not account for the size of the magnetosheath or the star\footnote{The min function is used in Equation \ref{eq:deltaf} because the bow shock in principal (depending on the value of $r_{m}$) can be larger than the star and the maximum $\delta_{F}$ is -100$\%$ (the whole star is blocked out when the bow shock is completely optically thick).}. In models of the bow shock incorporating more geometric sophistication (e.g., \citealt{Llama2011}), the absorption due to the bow shock occurs primarily in a thin sliver at the leading edge of the shock (see fig. 2 in \citealt{Llama2011}). Our assumption, that the occulting area is that of the entire half circle, is an overestimation; however, this is preferable as the resulting transit depths can be viewed as upper limits with geometric constraints producing smaller values of $\delta_F$. In order to be compared to observations $\delta_{F}$ would need to be convolved with the filter bandpass or spectral resolution. 

\begin{figure}{\phantom H}
\center
\begin{tabular}{c}
\vspace{0.25cm}
\includegraphics[width=1\linewidth]{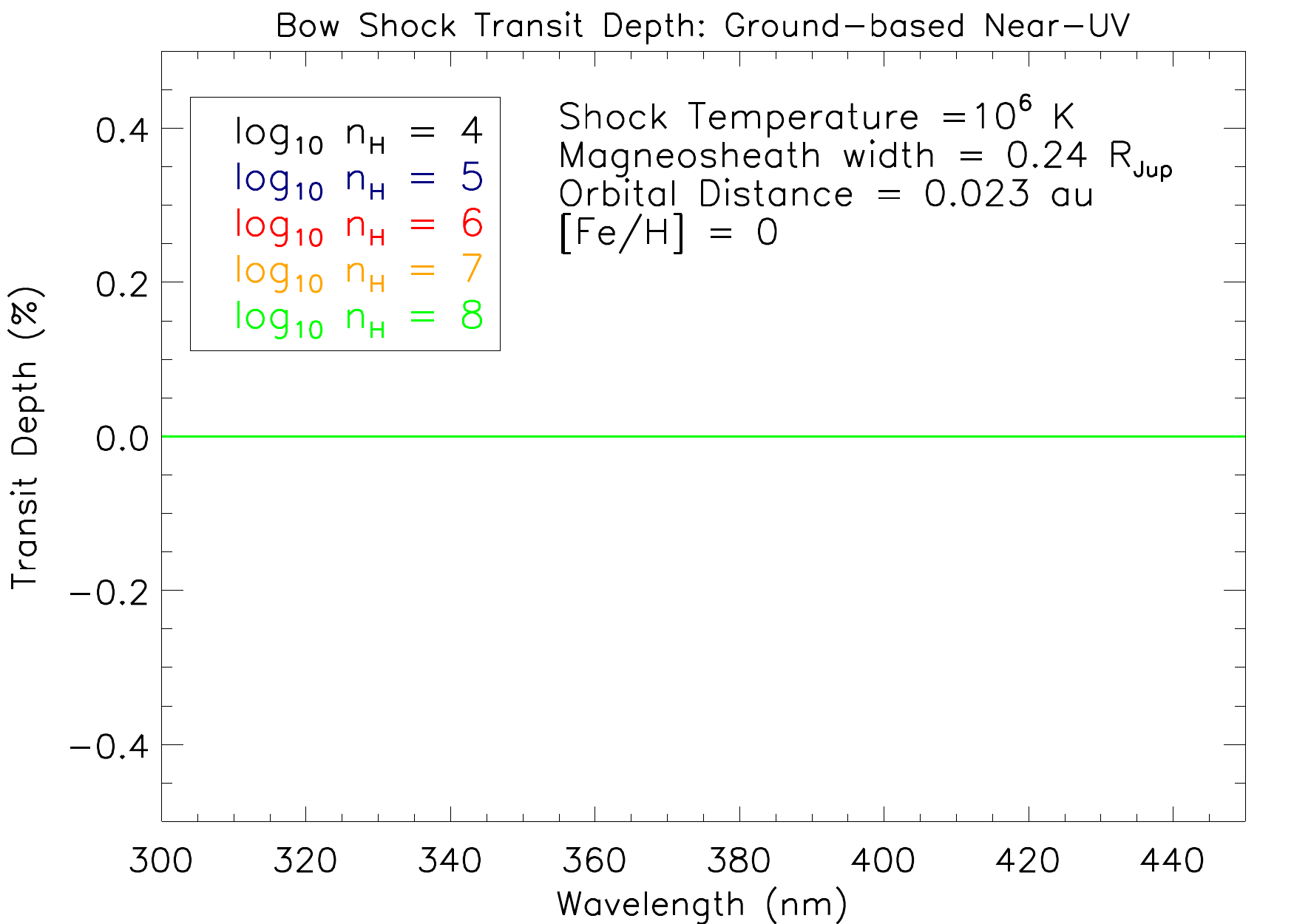}  \\
\vspace{0.25cm}
\includegraphics[width=1\linewidth]{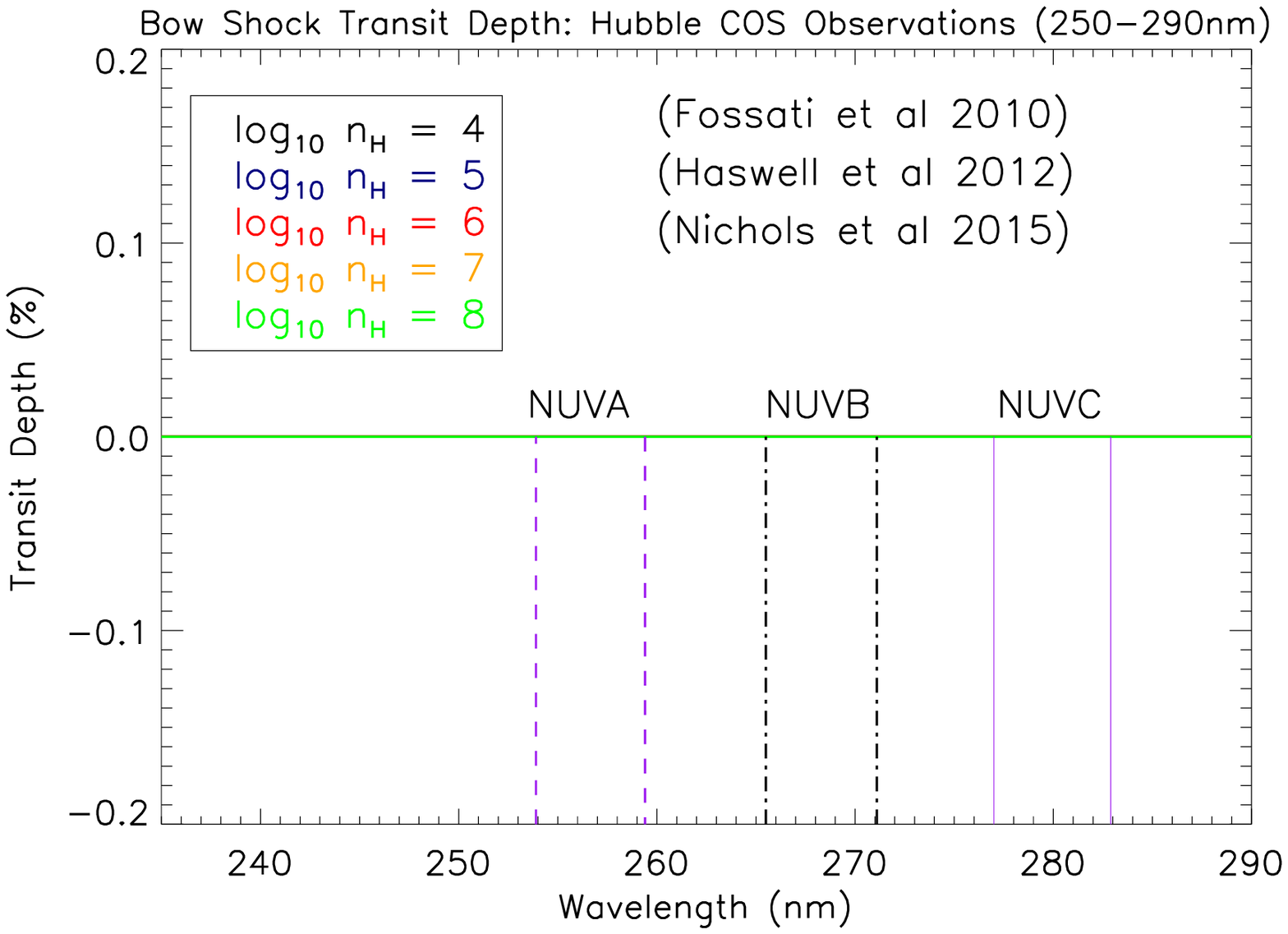}  \\
\includegraphics[width=1\linewidth]{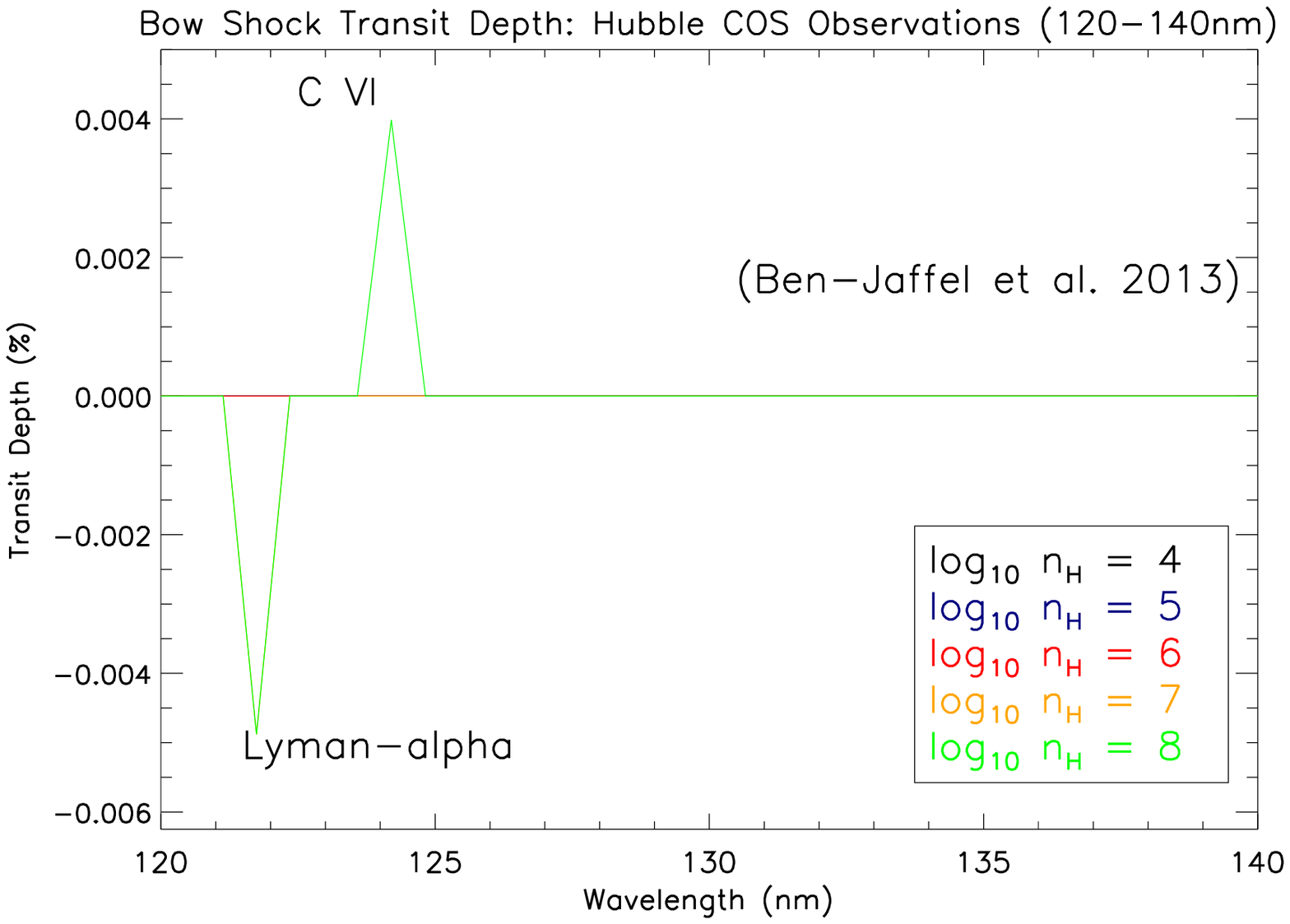}  
\end{tabular}
\caption{Transit depths due to the magnetosheath for the nominal model (Table \ref{tb:CLOUDY_modelsetup}) at varying hydrogen densities ($n_{H}$) for ground-based (\textbf{Top}) and \textit{Hubble Space Telescope} (\textbf{Middle/Bottom}) UV wavelengths. We do not detect any UV absorbing species under realistic conditions for the shocked coronal plasma that could cause an early UV ingress \citep{Vidotto2011a}. }
\label{fig:drop}
\end{figure}

The results of determining the transit depth due to the bow shock are in Figure \ref{fig:drop} for UV wavelengths. For the expected value $n_{\rm H} \simeq 10^4\ {\rm cm^{-3}}$ of the coronal density, and indeed even for $n_{\rm H}$ larger than the expected value by a factor of $10^3$, the transit depths are orders of magnitude too small to be observable. At the unphysically large density $n_{\rm H}=10^8\ {\rm cm^{-3}}$ expected at the base of the corona ($10^{-3}\ R_\star$) above the photosphere, a hydrogen Lyman-alpha absorption feature and a C VI {\it emission} feature are apparent. Both are still a factor $\sim 10^3$ too small to be measurable by \textit{HST}. Note that the emission feature from highly ionized C VI is in emission as the slab is brighter than the background star at such unphysically large columns for C VI in the slab.

In order to further interpret our results, we can estimate the optical depth of spectral lines. The optical depth of a spectral line would be 
\begin{align}
\tau = \int_{0}^{x} \sigma n(x) \ dx \label{eq:tau_int}
\end{align}
where $\sigma$ is the cross section, $n$ is the number density of that species, and $x$ is the width of the cloud. The cross section at line-center is equal to
\begin{equation}
\sigma = \frac{\sqrt{\pi} e^2}{m_{e} c} \frac{f \lambda_{0}}{v_{t}} \label{eq:sigma}
\end{equation}
where $e$ is the electric charge of an electron, $m_{e}$ is the mass of an electron, $c$ is the speed of light, $\lambda_{0}$ is the wavelength at line center, $v_{t} = \sqrt{2kT/m_{H}}$ is the thermal velocity, and $f$ is the oscillator strength. For illustration, the optical depth of Lyman-alpha at line-center is
\begin{align}
\tau_{Ly\alpha} =&\frac{\sqrt{\pi} e^2}{m_{e} c} \frac{f \lambda_{0}}{v_{t}} n_{1s} x \label{eq:tau_num}\\
\tau_{Ly\alpha} =&1.0 \left( \frac{n_{H}}{1.98\times10^{11} \ {\rm cm}^{-3}} \right), \label{eq:tau_ly}
\end{align}
where $n_{H}$ is the density of hydrogen nuclei (protons or atoms or molecules) and the density of hydrogen in the 1s state is $n_{1s} = 10^{-6.589} n_{H} \ {\rm cm}^{-3}$ (Table \ref{tb:CLOUDY_ion}). To determine the pre-factor in Equation \ref{eq:tau_ly}, the values of the variables we use in Equation \ref{eq:tau_num} are $\lambda_{0} = 121.56701 \ {\rm nm}$, T = $10^{6}\ {\rm K}$, $f$ = 0.4164 \citep{NIST_ASD}, and x = $2\Delta r_{m}$ (Table \ref{tb:CLOUDY_modelsetup}). Therefore, for a $n_{H} = 10^{8} \ {\rm cm}^{-3}$ we find a $\tau = 0.0005$ which is consistent with the value found from \texttt{CLOUDY} of 0.0001 and the transit depth in Figure \ref{fig:drop}. Lyman-alpha is very optically thin, which is why we see little absorption and no absorption for $n_{H}$ below $10^{8} \ {\rm cm}^{-3}$.

The cloud optical depth is used from the \texttt{CLOUDY} modeling to determine if absorption occurs. The optical depth of the cloud is calculated to be between $\sim 2.664\times10^{-11}$ and $2.663\times10^{-8}$ for all UV and optical wavelengths for densities between $10^{4}-10^{7} {\rm cm}^{-3}$, respectively. For a density of $10^{8} {\rm cm}^{-3}$ the optical depths are calculated to be $2.663\times10^{-7}$ across the UV and optical except for the Lyman-alpha and C VI line. All of these optical depths are too small to cause any detectable absorption (Figures \ref{fig:spec} and \ref{fig:drop}). 

Additionally, we perform a thorough parameter search with \texttt{CLOUDY} to determine what conditions can cause absorption in the UV and to check the robustness and biases of the nominal parameters (Table \ref{tb:CLOUDY_modelsetup}). The following cloud parameters are explored:
\begin{enumerate}
\item Temperature from 2,000 to $2\times10^{6}$ K
\item $[Fe/H]$ from 0 (solar) to +1
\item $\Delta r_{m}$ from 0.24 $R_{Jup}$ to 3 $R_{Jup}$
\item $n_{{\rm H}}$ values from $10^{4}\ {\rm cm^{-3}}$ to  $10^{12} \ {\rm cm^{-3}}$
\item Stellar luminosity from 1 to 5 $L_{\sun}$
\end{enumerate}
The results are presented in Figure \ref{fig:other}. The top panel shows the result of scaling the coronal density up to levels a factor $\sim 10^8$ above that expected at the orbital radius of the planet, and a factor of $10^4$ higher than that expected at the base of the corona. 
Even at such high densities, transit depths over the range $300-450 {\rm nm}$ are still at the $< 0.1\% $ level.
 The middle panel of Figure \ref{fig:other} shows the result of using an unphysically low value for the shock temperature ($10^4\ {\rm K}$) of the coronal gas. In this case, absorption lines for H I, Si II, and C II exhibit transit depths between $\sim 0.2 - 1.5 \%$ at very high densities $10^7-10^8\ {\rm cm^{-3}}$. Lastly, the bottom panel of Figure \ref{fig:other} again shows transit depths for a shock temperature $10^4\ {\rm K}$, now in the NUV (NUVA, NUVB, NUVC) bands.  Again at unphysically high densities there is now absorption by He II, Mn II, Mg I, and Mg II. The results of the bottom panel are at odds with the detection of Mg I but no Mg II absorption in HD 209458b, a typical hot Jupiter \citep{VidalMadjar2013}. The mean ionization fraction averaged over the width of the cloud (T=$10^{4}$ K) for Mg I = $10^{-13.092}$ and Mg II = $10^{-9.181}$. Therefore, this result suggests that the temperature of the planetary gas producing the Mg I absorption in HD 209458b is at a lower temperature, consistent with the temperature upper limits found by \citet{VidalMadjar2013}.

\begin{table}
\caption{Mean ionization fractions for the coronal gas for the ionization states of interest for H, He, C, Ne, Na, Mg, Al, Cl, Ca Sc, Cr, Mn, Fe, Co, Ni, Nu, and Cu calculated from the \texttt{CLOUDY} modeling }
\begin{tabular}{cc|cc}
\hline
\multicolumn{1}{c}{ State} &
\multicolumn{1}{c}{Fraction ($10^{X}$)} &
\multicolumn{1}{c}{ State} &
\multicolumn{1}{c}{Fraction ($10^{X}$)} \\
\hline
H I     &-6.589        &H II &  $-1.119\times10^{-7}$\\
He II   &-4.535         &He III& $-1.267\times10^{-5}$\\
C I     &-16.531	    &	C II &	-11.258 \\
C VI    &-0.243        & Ne I    &19.103\\
Ne II    &-14.348       & Ne III  &-10.478 \\
Ne IX    &-0.036       & Na I  &-20.064  \\
Na II    &13.898       & Na X     &-0.175\\
Mg I  &-19.528          &Mg II &-14.041  \\
Mg IX &-0.483          &Al I  &-19.849 \\
Al II & -14.529        &Al VIII &-0.399\\
 Cl I  &-23.207        &Cl II   &-17.431\\
 Cl III& -12.58        &Cl VIII   &-0.346\\
 Ca I  &-24.387       &Ca II   &-17.916\\
 Ca III& -13.829      & Ca XI       &-0.064\\
Sc II &-19.240         &Sc XII &-0.149  \\
Cr I  & -27.072         &Cr II &-20.507 \\
Cr X&    -0.336        & Mn I   &-27.606        \\
Mn II  &-21.056        &Mn X  &-0.328\\
Fe I   & -25.168       &Fe II   &19.027\\
Fe III & -14.267       &Fe IX  &-0.395\\      
Co I  &-28.678         &Co II   &-22.135 \\
Co X   &-0.659         &Ni I   & -27.872         \\ 
Ni II  &-21.656        &Ni XI  &-0.374  \\
Cu I   &-27.100        &Cu II  &-21.052          \\
Cu III & -15.717     &Cu X &-21.052\\
\hline
\end{tabular}
\vspace{-2em}
\tablecomments{The values in this table were taken for the nominal \\
model presented in Table \ref{tb:CLOUDY_modelsetup} with a hydrogen density of \\ $10^{8} \ {\rm cm^{-3}}$. Every \texttt{CLOUDY} model produced very similar trends.\\
The ionization fractions presented are averaged over the width \\of the cloud. }
\label{tb:CLOUDY_ion}
\end{table}

\subsection{Discussion} \label{sec:Discuss}
We find that using realistic parameters (Table \ref{tb:CLOUDY_modelsetup}) for the shocked coronal plasma in the \texttt{CLOUDY} modeling clearly indicates there is no species present that can absorb in the UV with any detectable optical depth (Figures \ref{fig:spec} and \ref{fig:drop}). This is also true for all other wavelengths examined (including optical). The mean ionization fractions of H, He, C, Ne, Na, Mg, Al, Cl, Ca Sc, Cr, Mn, Fe, Co, Ni, Nu, and Cu found using \texttt{CLOUDY} indicate that there aren't any spectral lines observed in the UV (e.g. Na I, Al I, Sc II, Mn II, Fe I, Co I, Mg I, Mg II) because the atoms are highly ionized (Table \ref{tb:CLOUDY_ion}). The results from our \texttt{CLOUDY} models appear to be at odds with the previous investigations of an early UV ingress  \citep{Vidotto2010,Vidotto2011a,Vidotto2011b,Vidotto2011c,Llama2011,llama2013}. The models in these papers suppose the source of opacity is Mg II; however, at temperatures characteristic of the corona, magnesium is highly ionized with very little Mg II remaining (see Table \ref{tb:CLOUDY_ion}, also \citealt{2014ApJ...785L..30B}). While much of the literature focuses on Mg II, this applies more broadly: coronal gas, either pre- or post-shock, will consist primarily of highly-ionized species with no expectation of opacity from neutral or singly-ionized species. For example, there should be no expectation of Lyman or Balmer absorption coming from neutral hydrogen within the post-shock coronal gas or lines from lower energy levels such as Mg I and Fe I (\citealt{Haswell2012}; \citealt{Bourrier2013}). 

Interactions between the post-shock stellar wind gas and the planetary gas can result in a hot population of neutral hydrogen through charge-exchange reactions which is potentially observable (e.g., \citealt{Tremblin2013}); however, this is not an observation of the shocked stellar wind gas, as the hot population is generated downstream at the interaction layer (contact discontinuity) between the post-shocked gas and the planetary gas (e.g. \citealt{Kislyakova2014}). For a more detailed discussion, see \S \ref{sec:planetary_stellar}.

\begin{figure}
\center
\begin{tabular}{c}
\vspace{0.25cm}
\includegraphics[width=1\linewidth]{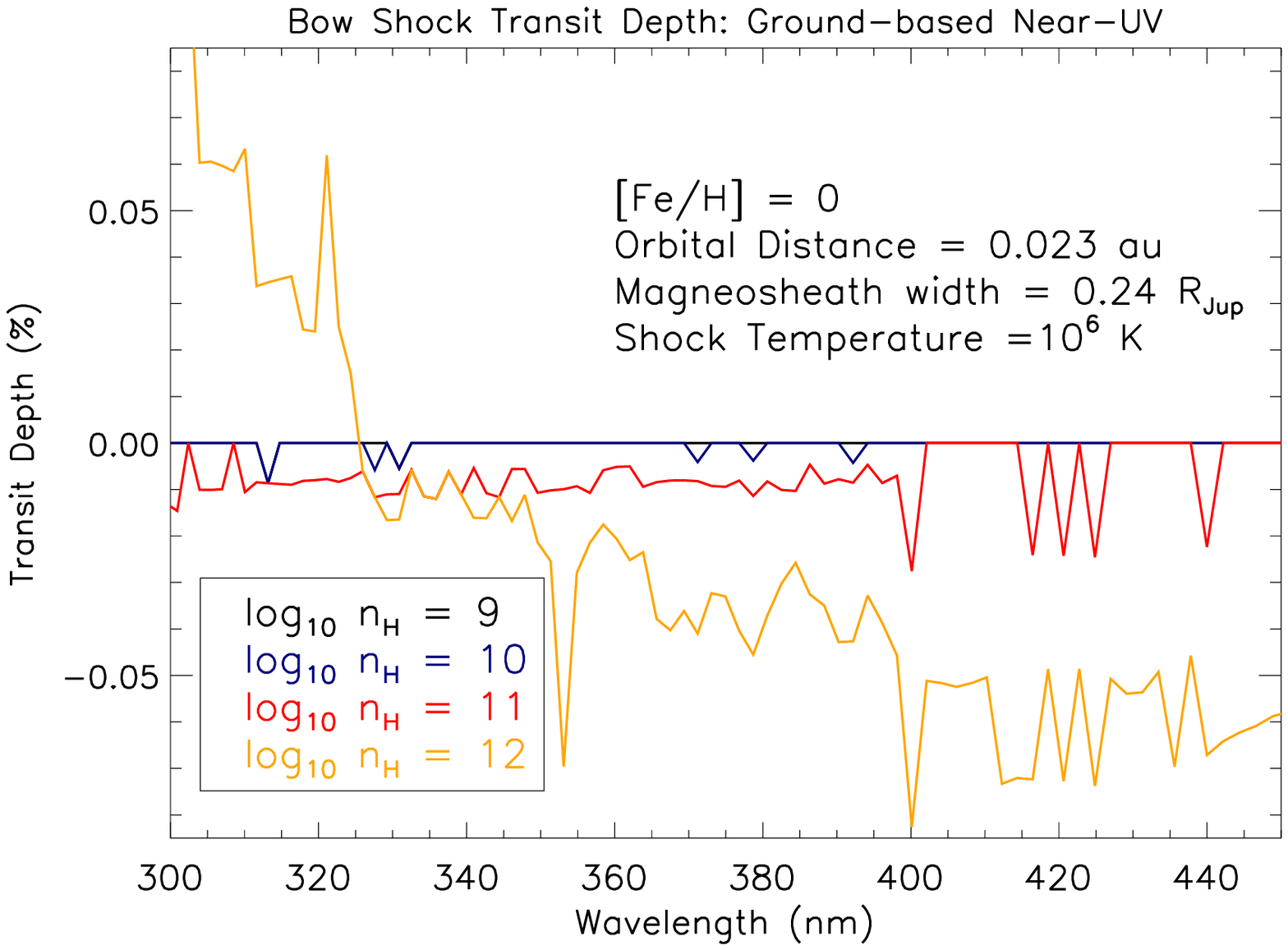}  \\
\vspace{0.25cm}
\includegraphics[width=1\linewidth]{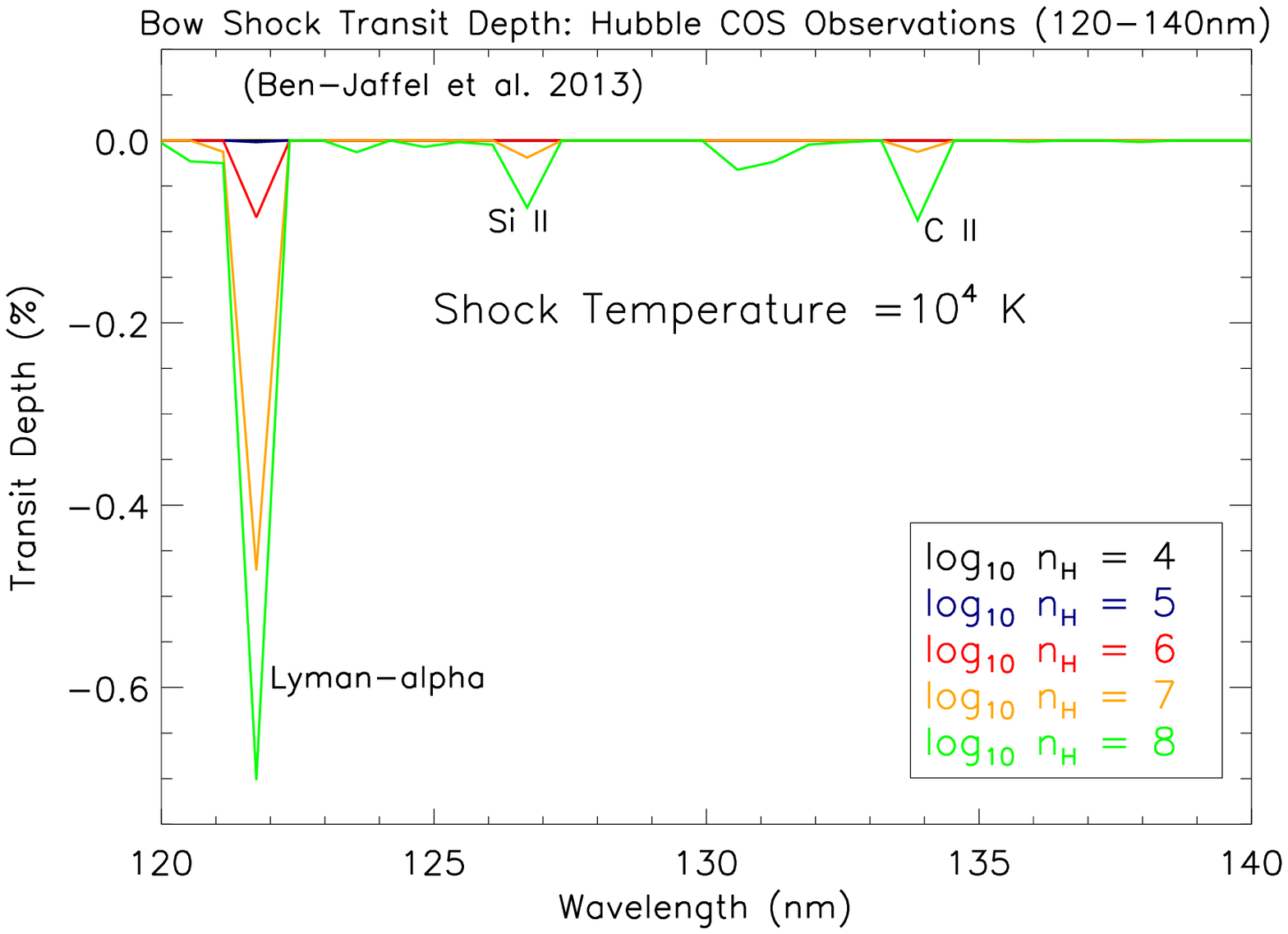}  \\
\includegraphics[width=1\linewidth]{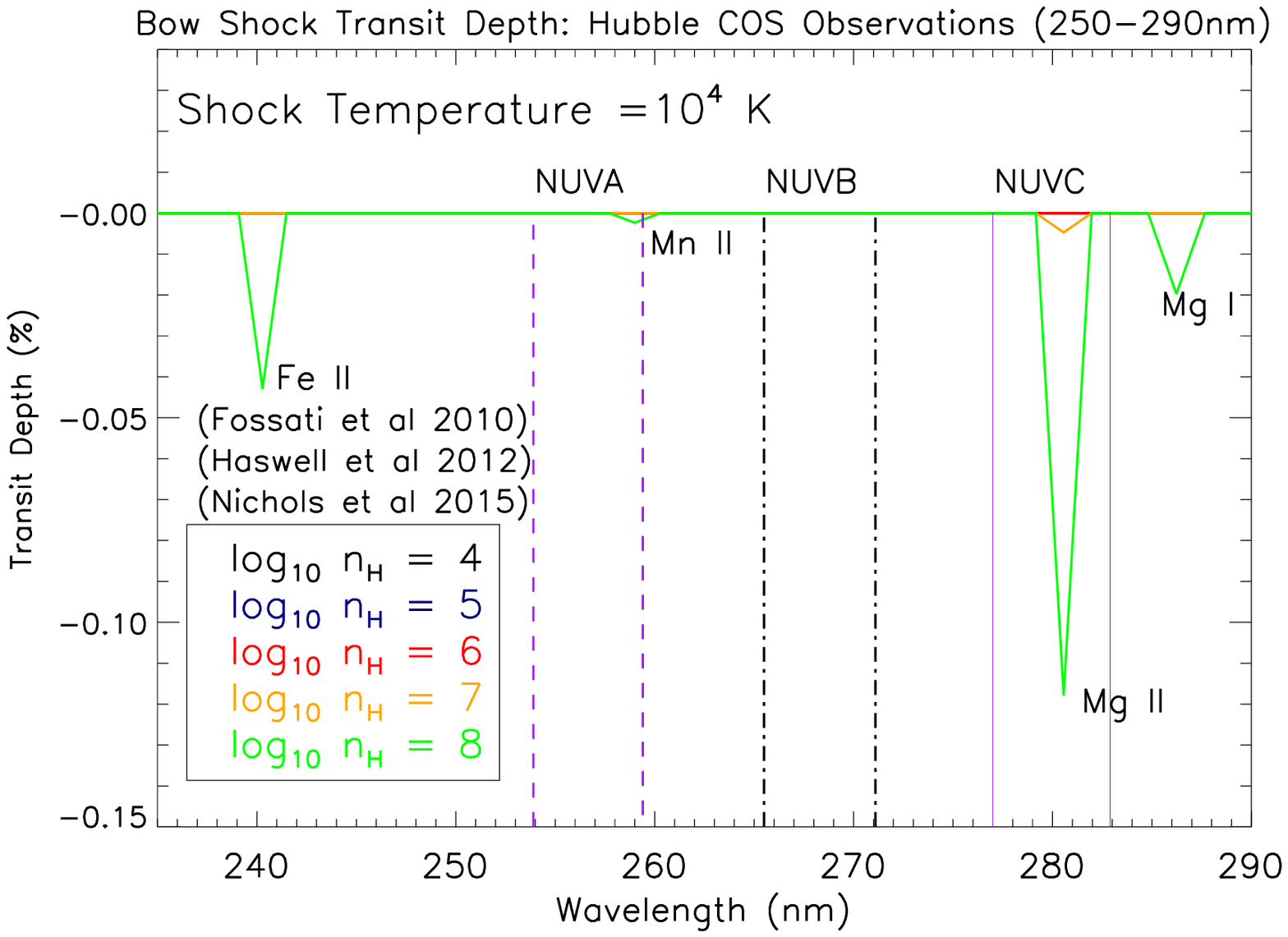}  
\end{tabular}
\caption{Transit depths due to the magnetosheath for \textbf{Top:} ground-based near-UV wavelengths for above $n_{H}= 10^{10} \ {\rm cm^{-3}}$, \textbf{Middle/Bottom:} HST UV wavelengths with the nominal model except the cloud temperature is 10,000 K.  }
\label{fig:other}
\end{figure}
 
From the parameter search, we determine the robustness of our results and under what conditions UV absorption can actually occur. We find for reasonable shock temperatures between $8\times10^{5}$ -- $2\times10^{6}$ K (consistent with the coronal temperatures measured for F-, G-, K-, and M-stars: \citealt{Aschwanden2005}; \citealt{Vaiana1981}) that the absence of an UV absorbing species did not change. Additionally, changing the $[Fe/H]$, $\Delta r_{m}$, and the stellar luminosity for the nominal model (Table \ref{tb:CLOUDY_modelsetup}) did not help with detectability. Therefore, our result of not finding any UV absorbing species is robust. In order for the C II 133.5 nm line to appear (as in the \citealt{Jaffel2013} HD 189733b observations) we need a plasma temperature between 2,000-100,000 K (Figure \ref{fig:other}) suggesting that this absorption is not coming from the stellar corona but likely from a gas with a much lower temperature such as an escaping planetary atmosphere. We find that no lines or continuum absorption sources appear in the NUV (NUVA, NUVB, NUVC filters) space-based wavelengths unless the cloud temperatures are below 10,000 K  (Figure \ref{fig:other}). This suggests that whatever caused the early ingress in WASP-12b (\citealt{Fossati2010b}; \citealt{Haswell2012}; \citealt{Nichols2015}) was not from shocked stellar wind gas but is more likely planetary gas as has been previously suggested (\citealt{Lanza2009}; \citealt{Lai2010}; \citealt{Bisikalo2013}; \citealt{Bisikalo2013A}; \citealt{Cherenkov2014A}; \citealt{Lanza2015}). An escaping atmosphere of WASP-12b would also explain the variability in the complete set of near-UV observations (\citealt{Fossati2010b}; \citealt{Haswell2012}; \citealt{Sing2013};  \citealt{Nichols2015}). For the case of a hot ($T \sim 10^6\, {\rm K}$) gas, we find that significant opacity (mainly metal lines) only appears for the ground-based near-UV wavelengths under \textit{unrealistic} conditions of a $n_{H}$ above $10^{10} \ {\rm cm^{-3}}$ for the nominal model (Figure \ref{fig:other}). Also, our modeling shows that for a density of $10^{12} \ {\rm cm^{-3}}$ the cloud actually becomes brighter than the star at wavelengths lower than 320~${\rm nm}$.\\
\indent Due to our results, any interpretation of UV absorption which relies on shocked coronal gas as the source of opacity  (\citealt{Vidotto2010}; \citealt{Turner2013a}; \citealt{Pearson2014}; Turner et al. in press) may need to be re-evaluated. This result also applies more broadly to all wavelengths, including interpreting optical asymmetry observations with magnetic bow shocks models. We believe this result is robust due to the fact that we covered the likely parameter space and found no effect. Our conclusions add to a body of theoretical work suggesting that UV asymmetry observations are not suitable approach for exoplanet magnetic field detection (\citealt{2014ApJ...785L..30B}; \citealt{G2015}; \citealt{Alexander2015}). Additionally, our modeling of lower temperature clouds suggests that the UV and optical observations are likely caused by gas from the planetary atmosphere (\citealt{Lai2010}; \citealt{Bisikalo2013}; \citealt{Bisikalo2013A}; \citealt{Cherenkov2014A}). Therefore, any future models attempting to explain the early UV observation need to include planetary gas in their simulations. The modeling described here does not provide any additional constraints on whether a bow shock exists around close-in exoplanets (see \citealt{Saur2013} and VJH11a for conflicting arguments) but does show that the bow shock does not produce any observable signature in the UV and optical.\\
\indent We, however, acknowledge that improvements to our modeling could be considered, such as using a more realistic density structure in the magnetosheath and stellar wind profile, including (magneto-) hydrodynamics (e.g. \texttt{TPCI}; \citealt{Salz2015}), and 3-D simulations (using \texttt{pyCloudy}; \citealt{Morisset2013}). Given the broad range of parameters tested here, we do not expect such improvements to change our conclusions.

\section{Modeling of planetary gas with \texttt{CLOUDY}} \label{sec:planetary}

Next, we simulate planetary gas in thermal and ionization equilibrium with the radiation field. The \texttt{CLOUDY} model transmission spectra presented in this paper for the planetary gas allow for a comprehensive study of 30 atomic elements and many molecules in ionization and thermal equilibrium with the stellar and diffuse radiation field. Since \texttt{CLOUDY} balances all the radiative and collisional rates self-consistently, the present study is able to make predictions for a number of transitions which have not yet been observed for any exoplanet.

We ran \texttt{CLOUDY} with the same exact geometry setup as in \S~\ref{sec:coronal} (Figures \ref{fig:cloudy_geo}), a width of $1 \,{\rm R_{Jup}}$, a covering factor of 0.0041, geometric factor in Equation \ref{eq:deltaf} of 0.401 (40.1$\%$), an orbital distance of $0.023\, {\rm au}$, and a solar metallicity for the cloud. The external radiation field was the same as used in \S~\ref{sec:coronal} (Figure \ref{fig:solar}). The hydrogen density ($n_{H}$) in the cloud is set to $10^{7}$ and $10^{9} \, {\rm cm^{-3}}$. The former choice is representative of the planetary gas found a few planetary radii from the planet (e.g. \citealt{Koskinen2014}). The latter choice is where the gas becomes opaque to Lyman continuum photons (\citealt{Clay2009}) and could be caused by escaping gas in a thick column (e.g. Roche lobe overflow; \citealt{Lai2010}). The gas kinetic temperature for this model is set by balancing the heating and cooling in the gas and the temperatures calculated by \texttt{CLOUDY} for the $10^{7}$ and $10^{9} \, {\rm cm^{-3}}$ clouds are approximately 8725 and 12859 ${\rm K}$, respectively.

A brief discussion on the the limitations of our modeling for the escaping planetary gas is as follows. We assume solar metallicity for both models; however, this is a source of uncertainty. Atomic species which reach the atomic layer will be carried along with the EUV-driven wind; however, the degree to which these atmospheres remain well mixed up to $1\, {\rm \mu bar}$ pressures where the gas transitions from molecular to atomic is unknown \citep{Koskinen2014}. Since the mixing of the atmosphere is beyond the scope of this paper and goal is only to find potential sources of opacity, we view the assumption of solar metallicity to be sufficient. The use of the static slab geometry will introduce uncertainties in the amount of absorption as it fails to capture the density and velocity structures.  The denser parts of the atmosphere are likely not well characterized. This exercise does, however, still offer potential opacity sources to look for during future transit observations.

\subsection{Results and Discussion} \label{sec:planetary_results}

\begin{figure*}
\center
\begin{tabular}{cc}
\vspace{0.25cm}
\includegraphics[width=0.5\linewidth]{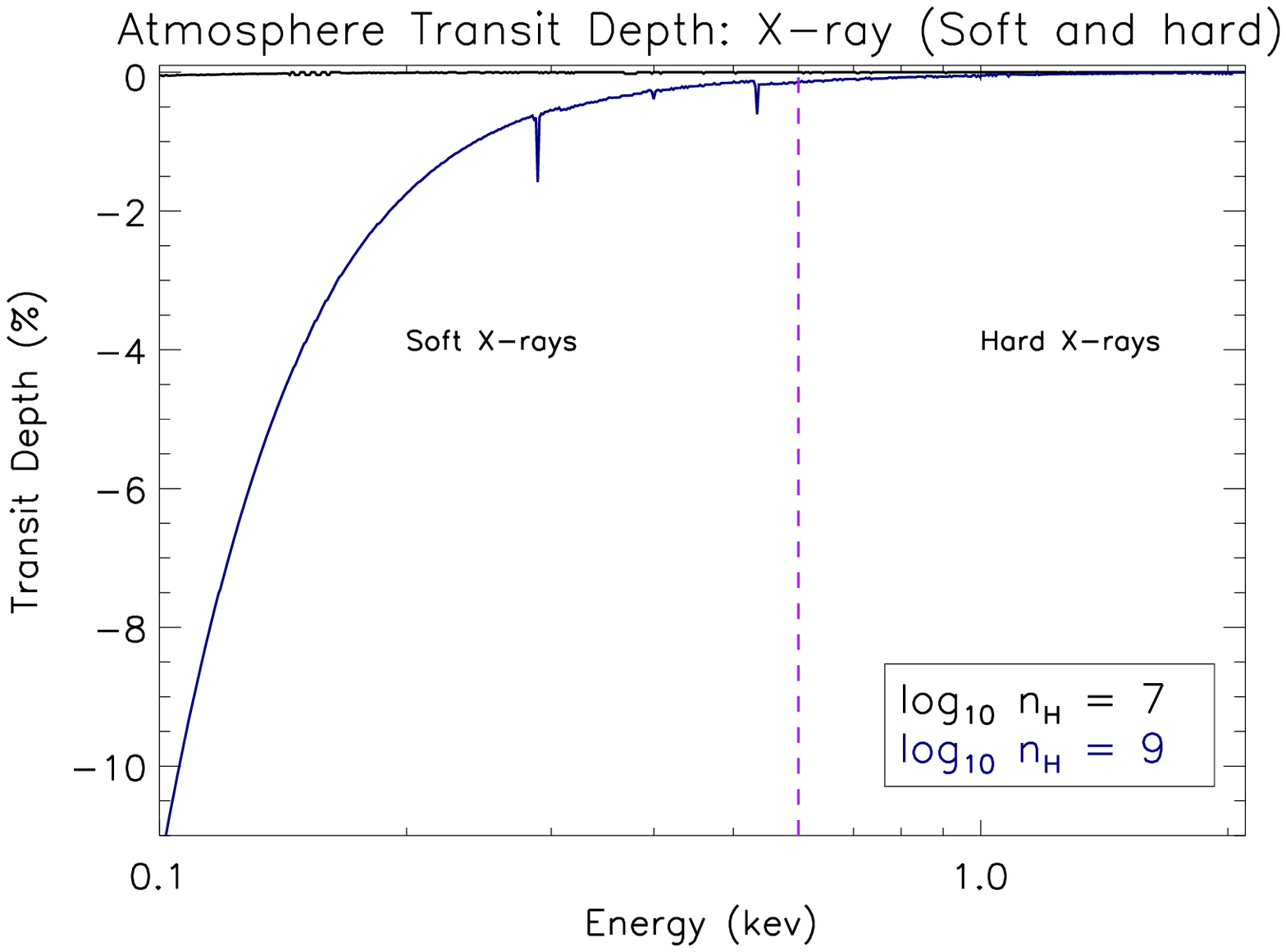}  &\includegraphics[width=0.5\linewidth]{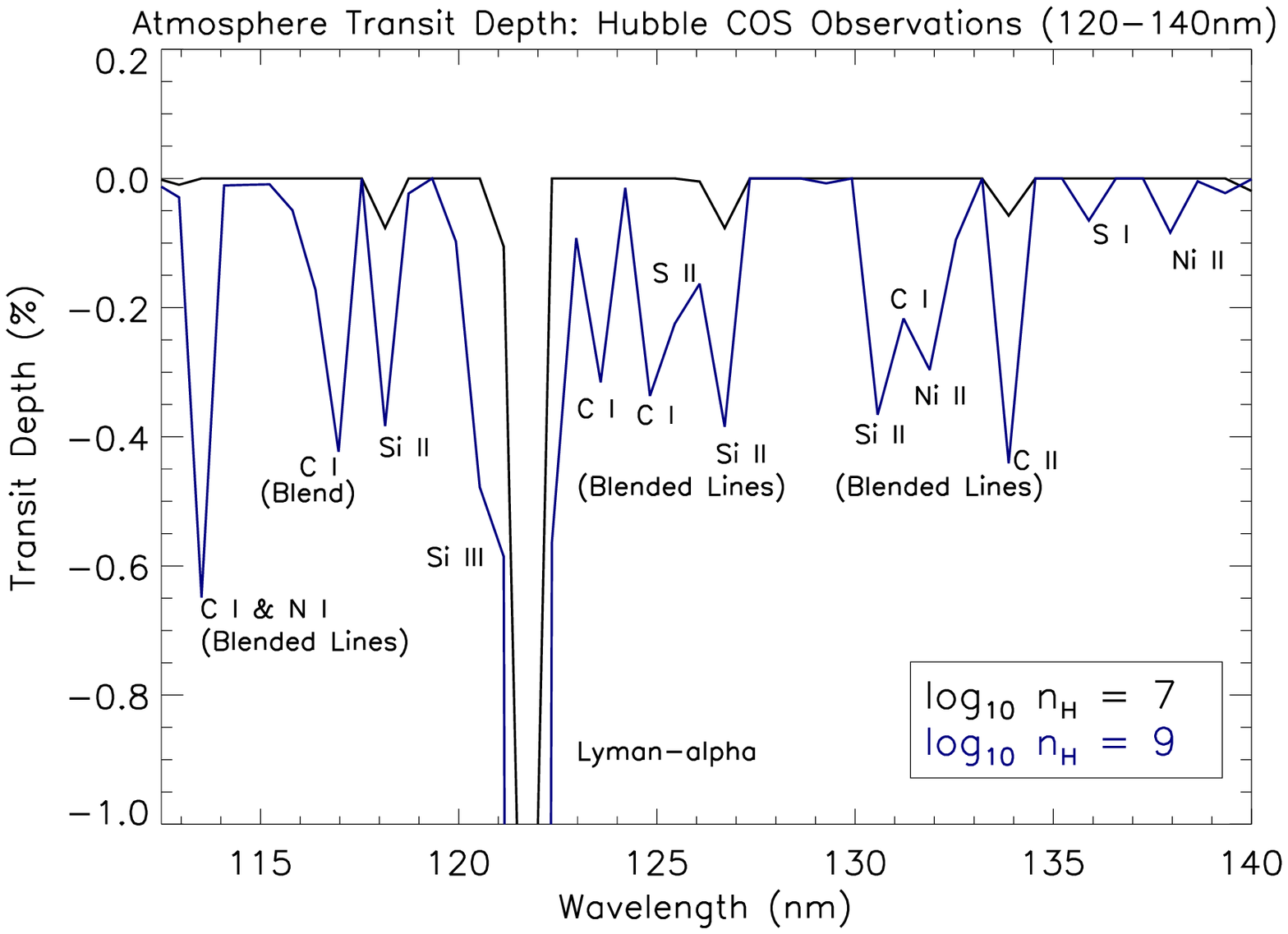}\\
\vspace{0.25cm}
\includegraphics[width=0.5\linewidth]{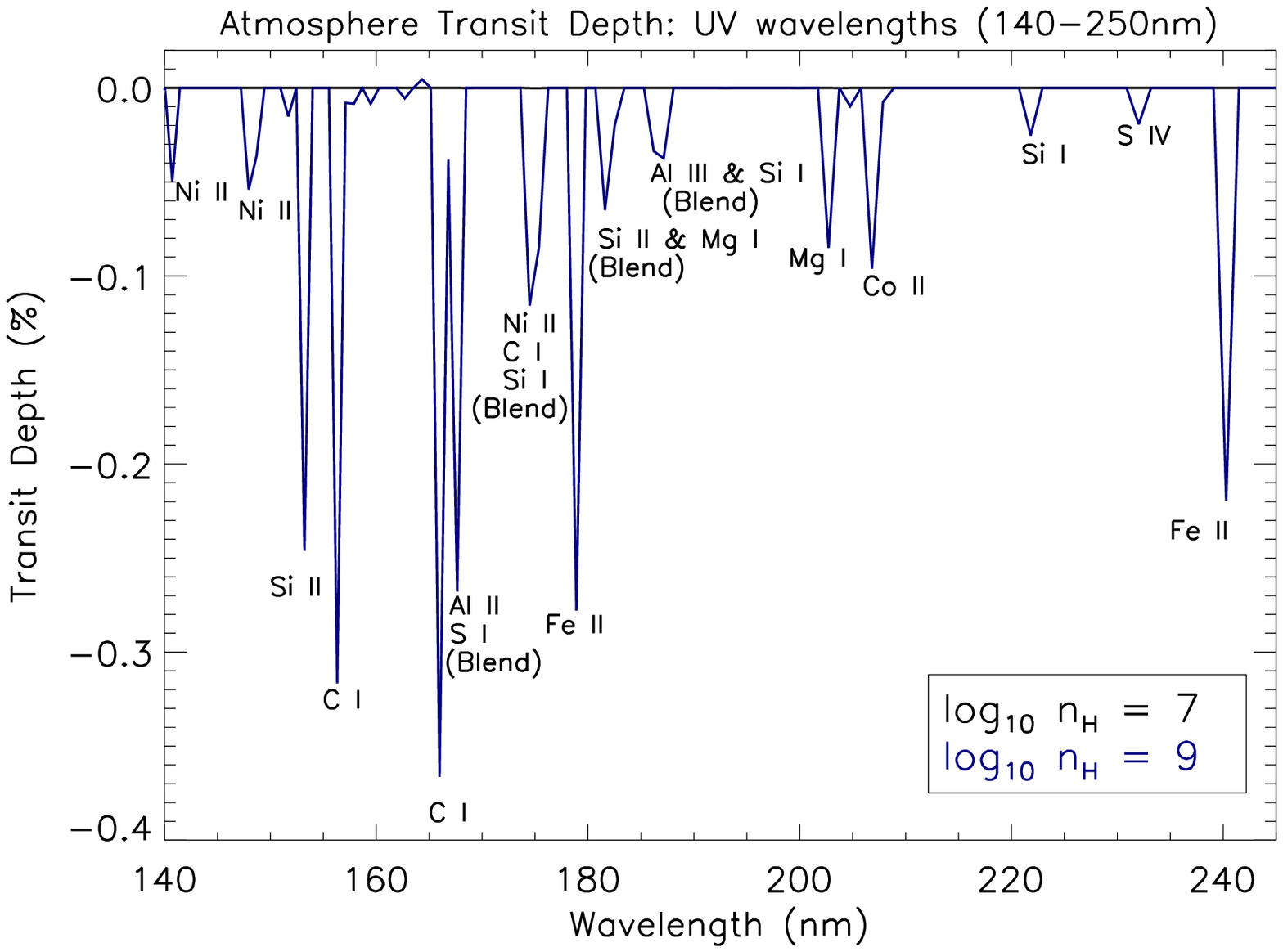}  &
\includegraphics[width=0.5\linewidth]{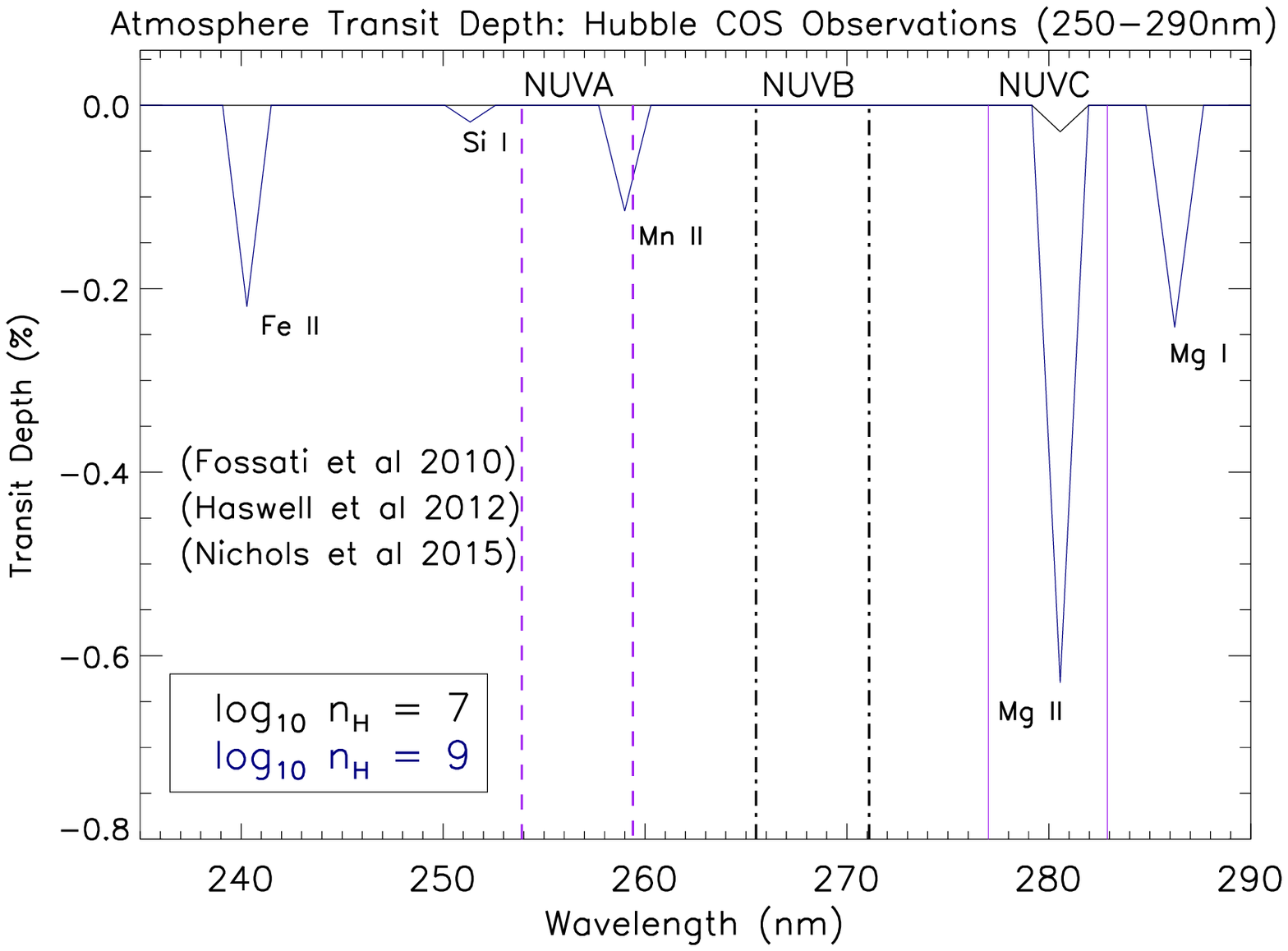}  \\
\includegraphics[width=0.5\linewidth]{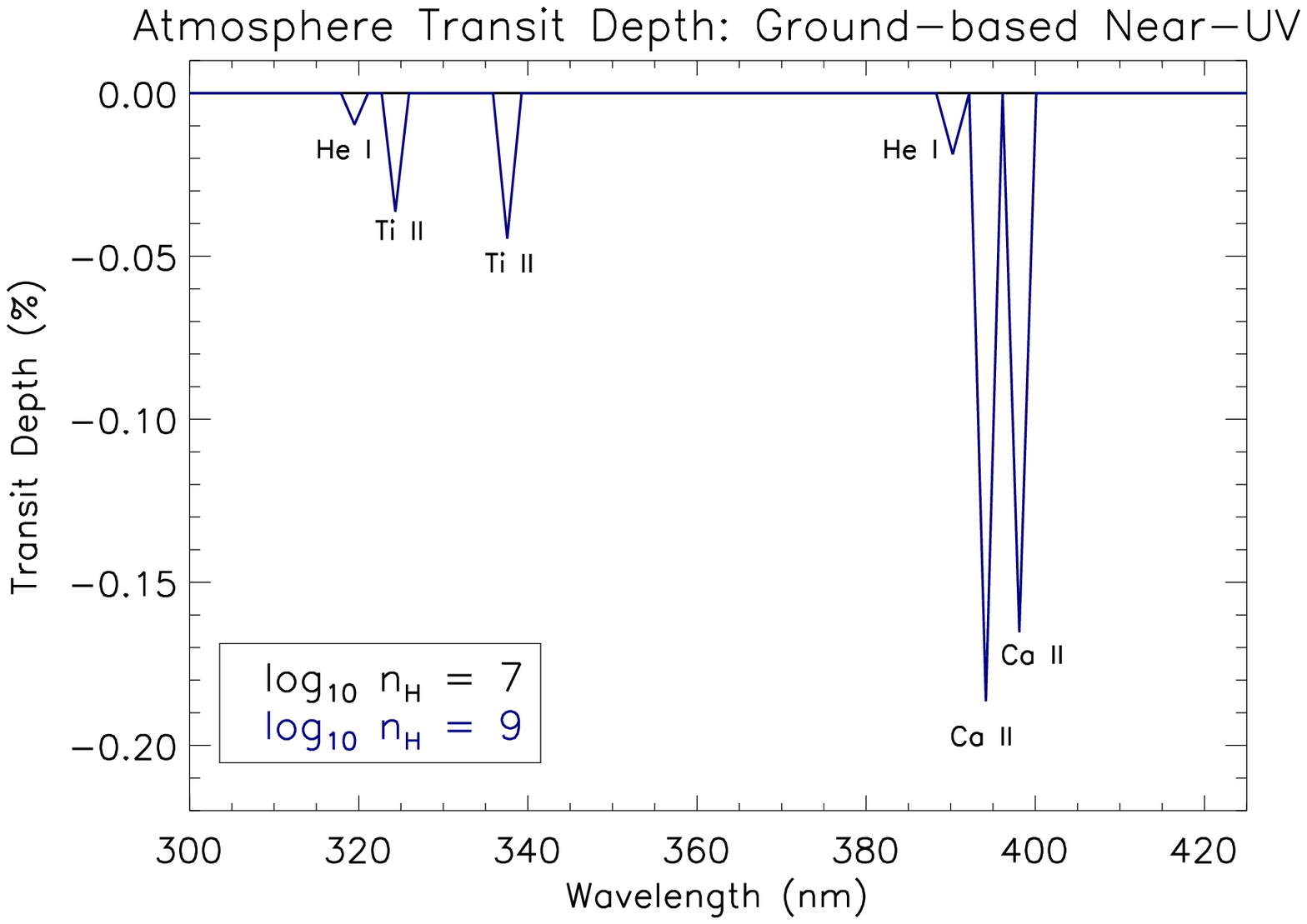}  &
\includegraphics[width=0.5\linewidth]{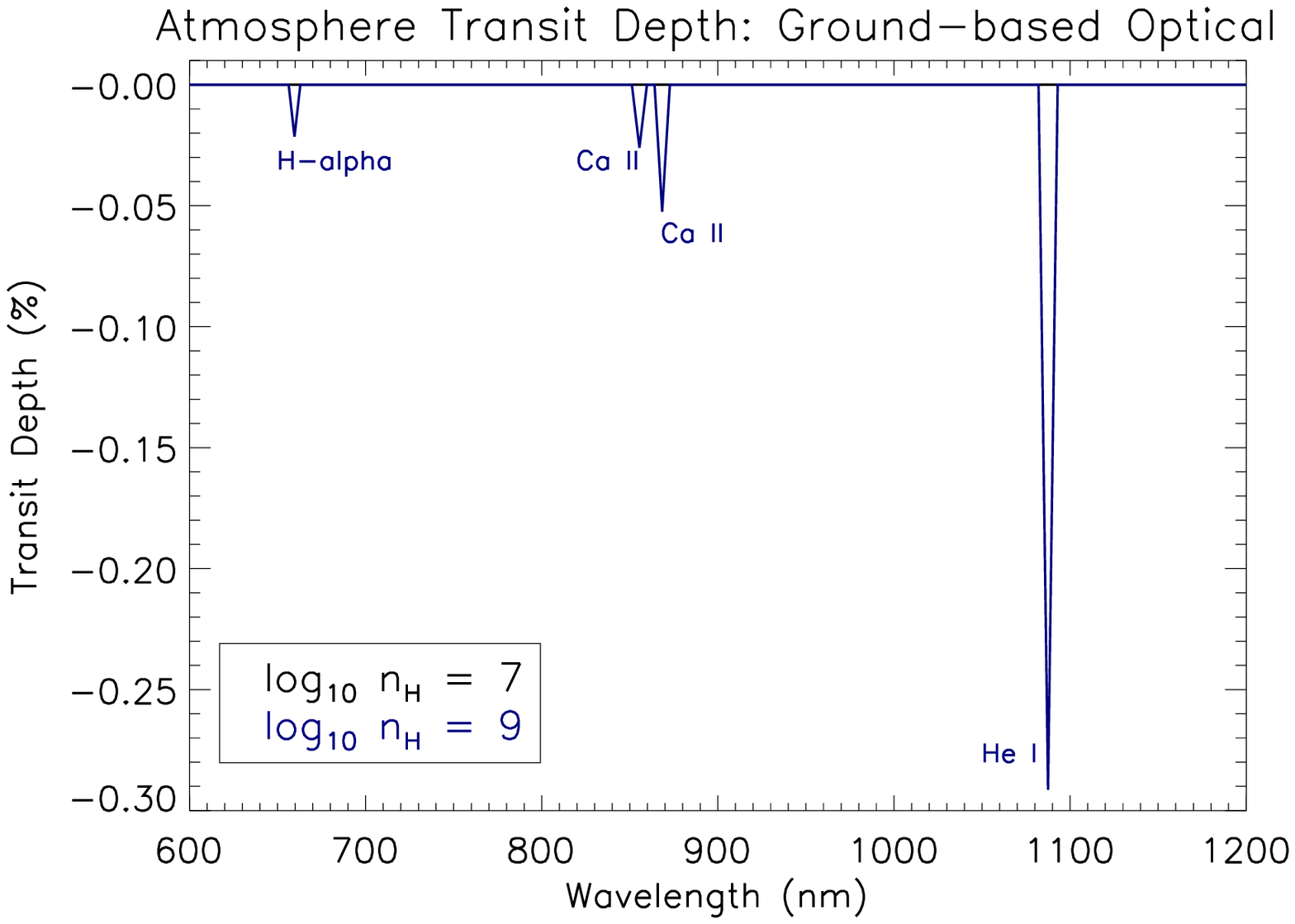} \\
\end{tabular}
\caption{Transit depths for the \texttt{CLOUDY} modeling of the escaped planetary gas in thermal equilibrium with the radiation field. We show the transit depth from X-rays to ground-based optical wavelengths. In the top-right plot, Lyman-alpha extends down to 12 $\%$. A complete list of all the lines predicted can be found in Table \ref{tb:spectral_predictions}.}
\label{fig:gas}
\end{figure*}

The output transit depths of the planetary gas are in Figure \ref{fig:gas} for X-ray, UV (space and ground), and optical and in Figure \ref{fig:gas2} for radio wavelengths. The mean ionization fractions of  H, He, C, N, Ni, O, Na, Mg, Al, Si, S, P, Ca, Ti, Mn, Fe, and Co found using \texttt{CLOUDY} are in Table \ref{tb:CLOUDY_ion_gas}. The most interesting results found for the planetary gas (Figure \ref{fig:gas}) are discussed below. The X-ray results (top-left) for a density $n_{\rm H}=10^9\ {\rm cm^{-3}}$ are consistent with the 8$\%$ transit depth observed on HD 189733b at X-ray wavelengths using Chandra \citep{Poppenhaeger2013}. The main source of opacity for the X-rays is bound-free from hydrogen and helium and the spectral lines are blended metal lines. The space-based UV wavelengths from 113-290 ${\rm nm}$ are shown (top-right, middle-left, middle-right). The most interesting lines that are potentially observable (this limit is anything with a transit depth greater than 0.10$\%$) in this wavelength regime (113-290 ${\rm nm}$) are H I, C I, C II, N I, Ni II, Mg I, Mg II, Al II, Si II, S I, Mn II, Fe II, and Co II. We find that the Lyman-alpha line now has an optical depth of 124 (very optically thick), as compared to the optically thin (0.0001) case for the coronal equilibrium. In the ground-based near-UV (bottom-left) the most promising observable lines are Ca II, He I, and Ti II (all not yet observed). In the optical regime (bottom-right), H-alpha, He I, and Ca II are the most promising lines. The detection of H-alpha in our modeling is consistent with the observations of H-alpha in the atmospheres of HD 189733b \citep{Jensen2012} and HD 209458b \citep{Defru2013}. For the ALMA wavelength range (Figure \ref{fig:gas}), binning bands 3-5 together results in a transit depth of 0.27 $\%$. While this transit depth is possibly observable (see also \citealt{Selhorst2013}), the problem with these bands are the fact that the hosts stars are very faint. The main source of opacity for frequencies lower than 200 ${\rm GHz}$ is free-free and between 200-400 ${\rm GHz}$ overlapping molecular lines are the dominate source. 

There are approximately 20 species (H I, C II, O I, Na I, Mg I, Mg II, Al I, Al III, Si III, K I, Ca I, Sc II, V II, Mn I, Mn II, Fe I, Fe II, Co I, Sn I, Eu III, Yb II) with lines currently observed in exoplanet upper atmospheres (e.g. \citealt{Charbonneau2002}; \citealt{VidalMadjar2003}; \citealt{VidalMadjar2004}; \citealt{Sing2008}; \citealt{Fossati2010b}; \citealt{Linsky2010}, \citealt{Sing2011}; \citealt{Haswell2012}; \citealt{Jensen2012}; \citealt{Jaffel2013}; \citealt{Defru2013};  \citealt{VidalMadjar2013}; \citealt{Bourrier2013}). Many of these same species are found in the \texttt{CLOUDY} modeling but we also find species (e.g. He I, C I, Al II, Si I/II, S II, Ca II, Ti II, Ni II, Mn II) and lines (e.g. Al III, Fe II, Mn I) that have not been observed. A comprehensive list of all the lines predicted can be found in Table \ref{tb:spectral_predictions}. Therefore, future observations are encouraged to search for these lines but more detailed theoretical models are still needed to shed light into opacity sources in escaping atmospheres of exoplanets.   

\begin{table}
\caption{Mean ionization fractions for the planetary gas for the ionization states of interest for H, He, C, N, Ni, O, Na, Mg, Al, Si, S, P, Ca, Ti, Mn, Fe, and Co calculated from the \texttt{CLOUDY} modeling }
\begin{tabular}{cc|cc}
\hline
\multicolumn{1}{c}{ State} &
\multicolumn{1}{c}{Fraction ($10^{X}$)} &
\multicolumn{1}{c}{ State} &
\multicolumn{1}{c}{Fraction ($10^{X}$)} \\
\hline
H I  	&	-0.083	&	H II      	&	-0.761	\\
He I 	&	-0.206	&	He II     	&	-0.433	\\
C I 	&	-0.868	&	C II      	&	-0.072	\\
C III 	&	-1.76	&	N I	&	-0.115	\\
N II	&	-0.632	&	Ni I  	&	-0.115	\\
  Ni II 	&	-0.632	&	O  I  	&	-0.069	\\
  O II  	&	-0.836	&	Na I  	&	-2.868	\\
   Na II  	&	-0.11	&	Mg I 	&	-1.585	\\
   Mg II    	&	-0.086	&	Mg III 	&	-0.811	\\
 Al I  	&	-3.943	&	Al II 	&	-0.014	\\
Al III	&	-1.711	&	Al IV	&	-1.906	\\
 Si II 	&	-0.001	&	Si III 	&	-2.657	\\
 S I   	&	-1.423	&	S II   	&	-0.038	\\
 P I   	&	-3.856	&	P II   	&	-0.028	\\
 Ca I  	&	-3.543	&	Ca II  	&	-0.087	\\
  Ca III 	&	-0.741	&	Ti I	&	-4.049	\\
Ti II	&	-0.033	&	Mn I      	&	-3.981	\\
  Mn II 	&	-0.011	&	Mn III    	&	-1.599	\\
  Fe I	&	-4.177	&	Fe II   	&	-0.005	\\
 Fe III  	&	-1.937	&	Fe IV   	&	-5.469	\\
Co I	&	-1.135	&	Co II	&	-0.033	\\
\hline
\end{tabular}
\vspace{-2em}
\tablecomments{The values in this table were from the model with a \\hydrogen density of $10^{9} \ {\rm cm^{-3}}$. The ionization fractions \\ presented are averaged over the width of the cloud. The gas \\ kinetic temperature calculated by \texttt{CLOUDY} was $\sim$12859 ${\rm K}$ (\S \ref{sec:planetary}).} 
\label{tb:CLOUDY_ion_gas}
\end{table}

\subsection{Planetary gas interacting with the stellar wind}
\label{sec:planetary_stellar}
The gas modeled in \S~\ref{sec:planetary_results} is assumed to reside in the planet's upper atmosphere. While interaction with the stellar radiation field was included, collisions between particles in the planetary gas and the stellar wind were ignored. However, there are models for the hydrogen Lyman-alpha absorption during transit for HD 209458b that invoke charge exchange of atoms from the planet with solar wind protons to create fast moving neutral atoms \citep{2008Natur.451..970H, Tremblin2013, Christie2016}. In the limit of short mean free paths for the atoms in the stellar wind gas, the energetic atoms occupy a hydrodynamic mixing layer created by eddies at the interface of the two gases \citep{Tremblin2013}. For finite mean free paths \citep{2008Natur.451..970H}, as long as the mean free path of the atoms is not much larger than the magenetosheath width, some atoms can collide with the stellar wind protons and be entrained with the magnetosheath flow around the planet.

While we have argued that stellar wind gas near the bow shock region cannot provide enough opacity to cause observable transit depths, atoms {\it originating from the planet}, and entrained with the stellar wind flow through the magnetosheath could provide this source of opacity, as long as they are not ionized too quickly by the hot stellar wind gas. Hence the geometry assumed in the magnetic or non-magnetic bow shock models may indeed be valid, although a sufficient source of atoms from the planet must always be considered. As movement of charged particles through the planetary magnetosphere may only occur along field lines, movement of neutral particles across field lines is subject to strong ion-neutral drag forces \citep{2004Icar..170..167Y,2011ApJ...728..152T}. Open field lines carrying an outflow from the polar regions of the planet are another possible source of neutrals \citep{2014ApJ...788..161T}.

It is important to distinguish between absorbing atoms originating in the planet, as is assumed here, and in the stellar wind, as is assumed in VJH11a. The source matters even if the geometry is the same. The key point is that the distribution of planetary neutrals decreases with distance from the planet and there would be a far greater number of absorbing species toward the planet. Hence for planetary neutrals one might expect absorption not only near the bow shock, but also all along the line from the bow shock to the planet. This would be a significant change to the model used to interpret the observations as compared to assuming the absorption is solely from the stellar wind gas.

\section{Conclusions}
Using the plasma photoionization and microphysics code \texttt{CLOUDY} (\citealt{Ferland1998}; \citealt{Ferland2013}) we explore whether there is a UV absorbing species in the stellar wind that can cause an early UV ingress in the transits of close-in exoplanets due to the presence of a magnetic \citep{Vidotto2011a} or non-magnetic (\citealt{Lai2010}; \citealt{Bisikalo2013}; \citealt{Bisikalo2013A}) bow shock compressing the coronal plasma. We find under realistic physical conditions for the corona (Table \ref{tb:CLOUDY_modelsetup}; \S~\ref{sec:coronal}) that there aren't any species that can cause an absorption with sufficient opacity for all UV wavelengths and also for all other wavelengths between X-ray and radio (Figures \ref{fig:spec} and \ref{fig:drop}). A thorough parameter search with \texttt{CLOUDY} is performed to check the robustness and biases of the nominal parameters and we find that our conclusions are robust (\S~\ref{sec:results}). Therefore models cannot posit a distribution of neutrals only out in the stellar wind, but rather if absorbers are mixed into the stellar wind, there must be an even larger population toward the planet which may cause absorption all the way between the planet and the bow shock. In other words, the bow shock geometry model which has been used to infer planetary magnetic fields is an incomplete model as it does not confront the likely existence of absorbers between the bow shock and the planet. Previous detections of an early ingress in WASP-12b (\citealt{Fossati2010b}; \citealt{Haswell2012}) and HD 189733b (\citealt{Jaffel2013}; \citealt{Cauley2015}) are likely caused by the planetary upper atmosphere or a mixture of planetary and stellar wind material as suggested by our modeling of clouds with lower temperatures (\S \ref{sec:Discuss}; Figure \ref{fig:other}). Our conclusions are consistent with other studies suggesting that UV asymmetry observations are not a suitable approach for detecting exoplanet magnetic fields (\citealt{2014ApJ...785L..30B}; \citealt{G2015}; \citealt{Alexander2015}) and that suggest an escaping atmosphere (\citealt{Lai2010}; \citealt{Bisikalo2013}; \citealt{Bisikalo2013A}; \citealt{Cherenkov2014A}) as the cause of the UV observations. 

\begin{figure}
\center
\includegraphics[width=1\linewidth]{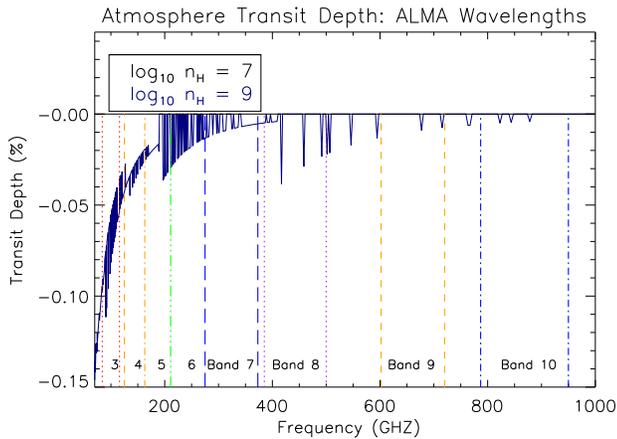} 
\caption{Transit depth for radio wavelengths for the \texttt{CLOUDY} modeling of the escaped planetary gas in thermal equilibrium with the radiation field.  }
\label{fig:gas2}
\end{figure}

We also simulate escaping planetary gas in ionization and thermal equilibrium with the stellar radiation field with \texttt{CLOUDY} (\S~\ref{sec:planetary}). From this modeling, we find species with strong absorption lines (Figure \ref{fig:gas} and \ref{fig:gas2}; Table \ref{tb:spectral_predictions}) previously observed in exoplanet upper atmospheres (e.g. \citealt{Charbonneau2002}; \citealt{VidalMadjar2003}; \citealt{VidalMadjar2004}; \citealt{Sing2008}; \citealt{Fossati2010b}; \citealt{Linsky2010}, \citealt{Sing2011}; \citealt{Haswell2012}; \citealt{Jensen2012}; \citealt{Jaffel2013}; \citealt{Defru2013};  \citealt{VidalMadjar2013}; \citealt{Bourrier2013}) but also make predictions for many species and lines that have not been observed (Table \ref{tb:spectral_predictions}; He I, C I, Al II, Si I/II, S II, Ca II, Ti II, Ni II, Mn II). Therefore, the \texttt{CLOUDY} modeling in this paper is a motivation for more detailed studies of possible absorbing species which may be observable in the transmission spectra of close-in exoplanets (see also, \citealt{Salz2015,Salz2015b}).

\begin{table*}
\caption{Spectral lines predicted for the planetary gas by \texttt{CLOUDY}.  }
\begin{tabular}{cccccccc}
\hline
\multicolumn{1}{c}{Vacuum (Air) $\lambda$} &
\multicolumn{1}{c}{Species} &
\multicolumn{1}{c}{Transit } &
\multicolumn{1}{c}{Previously} &
\multicolumn{1}{c}{Vacuum $\lambda$} &
\multicolumn{1}{c}{Species} &
\multicolumn{1}{c}{Transit } &
\multicolumn{1}{c}{Previously} \\
\multicolumn{1}{c}{[nm]} &
\multicolumn{1}{c}{} &
\multicolumn{1}{c}{Depth [$\%$] } &
\multicolumn{1}{c}{Observed} &
\multicolumn{1}{c}{[nm]} &
\multicolumn{1}{c}{} &
\multicolumn{1}{c}{Depth [$\%$] } &
\multicolumn{1}{c}{Observed} \\
\hline
1083.3306 (1083.303)  	&	He I         	&	0.28	& N	   	&	167.079	&	Al II                  		&	0.26 (blend)               	& N	\\
866.452   (866.214)  	&	Ca II        	&	0.052	&N	    	&	166.217	&	S I                    		&	0.26 (blend)             	&N	\\
854.444  (854.209)   	&	Ca II         	&	0.026	&N	    	&	165.7	&	C I                      		&	0.36	&	N\\
656.4614  (656.28)  	&	H-alpha        	&	0.021	&	Y (1)   	&	157.591	&	Co II                   		&	0.03	&	N\\
396.959  (396.847)  	&	Ca II          	&	0.16	&N		&	156.133	&	 C I                   		&	0.314	&	N\\
393.477  (393.366)  	&	Ca II          	&	0.19	&N		&	153.1	&	Si II                		&	0.24	&	N\\
388.9750 (388.865)  	&	He I           	&	0.019	&N		&	150	&	Fe II                  		&	0.015	&	N\\
336.571  (336.474)  	&	Ti II      	&	0.044	&	N	&	147.274	&	Ni II                    		&	0.0535	&	N\\
323.8078 (323.714)  	&	Ti II          	&	0.036	&N		&	140.037	&	Ni II                 		&	0.049	&	N\\
318.8667 (318.775)  	&	He I           	&	0.01	&N		&	137.573	&	Ni II                  		&	0.083	&	N\\
285.2965 (285.213)  	&	Mg I           	&	0.24	&	 Y (2)	&	135.605	&	S I                    		&	0.065	&	N\\
280.3531 (280.271)  	&	Mg II          	&	0.623	&	 Y (3)	&	133.5	&	C II                   		&	0.44	& Y (4)	\\
258.9746 (258.897) 	    &	Mn II           	&	0.11	&N	 	&	132.4117	&	Ni II                   		&	0.294 (blend)      	&N	\\
251.8226 (2517.47) 	    &	Si I            	&	0.01	&N	 	&	131.477	&	  C I                  		&	0.215 (blend)        	&N	\\
239.9997 (2399.27) 	    &	Fe II           	&	0.217	&N		&	130.766	&	  Si II                		&	0.362 (blend)        	&N	\\
233.5123 (233.441) 	    &	S IV            	&	0.0192 (blend)     	&	N 	&	126.332	&	Si II                   		&	0.381	&N	\\
233.5321 (233.46)  	    &	S IV            	&	 0.0192 (blend)    	&	 N	&	125.6	&	S II                   		&	0.162	&	N\\
221.500  (221.431)  	&	Si I            	&	0.025	&N		&	125.068	&	C I                     		&	0.223	&N	\\
206.156  (206.09)   	&	Co II           	&	0.095	&N	 	&	124.75	&	  C I                  		&	0.33	&	N\\
202.6477 (202.582) 	    &	Mg I            	&	0.84	&N		&	123.329	&	 C I                   		&	0.31	&	N\\
186.2789	           &	Al III                 	&	0.03 (blend)       	&	N	&	121.567	&	Lyman-alpha            		&	12.4 (blend)       	& Y (5)	\\
185.4716	            &	Al III                 	&	0.03 (blend)       	&	N	&	120.651	&	 Si III                		&	0.58 (blend)       	& Y	(6) \\
185.3047	&	Si 1                   	&	0.03 (blend)       	&	N	&	117.959	&	 Si II                 		&	0.38	&N	\\
181.399	&	Si II                  	&	0.2  (blend)       	&	N	&	116.681	&	 C I                   		&	0.42 (blend)      	&N	\\
181.7313	&	Mg I                   	&	0.08 (blend)       	&N		&	116.598	&	  C I                  		&	0.42 (blend)       	&N	\\
1786	&	Fe II                  	&	0.28	&	N	&	116.236	&	 C I                   		&	0.42 (blend)       	&N	\\
176.793	&	Si I                   	&	0.0843  (blend)           	&N		&	113.206	&	 C I                   		&	0.64 (blend)         	&N	\\
175.1823	&	C I                    	&	0.15  (blend)             	&N		&	113.112	&	 N I                   		&	0.64 (blend)       	&N	\\
174.424	&	Ni II                  	&	0.318  (blend)     	&N		&	113.046	&	 C I                    		&	0.64 (blend)       	&	N\\
\hline
\end{tabular}
\vspace{-2em}

\tablecomments{The model used has a hydrogen density of  $10^{9} \ {\rm cm^{-3}}$  and the gas kinetic temperature calculated by \texttt{CLOUDY} was $\sim$12859 ${\rm K}$ (\S \ref{sec:planetary}). Additionally, the default resolution model was not used to search for spectral lines due to many spectral lines overlapping each other. We used the line labels command (creates a list of all emission lines transported in the code) in \texttt{CLOUDY} to determine the species responsible for each spectral line in Figure \ref{fig:gas}.} 
\tablerefs{(1) \citealt{Jensen2012}; (2) \citealt{VidalMadjar2013}; (3) \citealt{Haswell2012}; (4) \citealt{Linsky2010}; (5) \citealt{VidalMadjar2003}; (6) \citealt{Bourrier2013}  }
\label{tb:spectral_predictions}
\end{table*}


\section*{Acknowledgments}

J. Turner and  R. Johnson were partially supported by the NASA's Planetary Atmospheres program. J. Turner was also partially funded by the Virginia Space Grant Consortium Graduate Research Fellowship Program and by the National Science Foundation Graduate Research Fellowship under Grant No. DGE-1315231. P. Arras and D. Christie were supported by NASA Origins of Solar Systems Grants NNX14AE16G and NNX10AH29G

We would like to thank Aline Vidotto, Moira Jardine, Christiane Helling, Joe Llama, and Andrew Collier Cameron for their insightful comments on this research. We thank Greg Ferland on his help with the \texttt{CLOUDY} modeling. \texttt{CHIANTI} was used in this research and is a collaborative project involving George Mason University, the University of Michigan (USA) and the University of Cambridge (UK). 

We would like to thank the anonymous referee for their useful comments which have led to considerable improvements in this paper.

\bibliographystyle{mn2e}
\bibliography{reference_new.bib}

\label{lastpage}
\end{document}